\documentclass[final,5p,times,twocolumn]{elsarticle}

\usepackage{graphicx}
\usepackage{color}
\usepackage{amsmath}
\usepackage{amssymb}
\usepackage{longtable,tabularx}
\usepackage{hyperref}
\usepackage{bm}
\usepackage{algorithm}
\usepackage{algpseudocode}
\usepackage[switch]{lineno}

\usepackage{xcolor}
\usepackage{tikz}
\usetikzlibrary{shapes.geometric}
\usetikzlibrary{arrows.meta}

\definecolor{tabblue}{HTML}{1F77B4}
\definecolor{taborange}{HTML}{FF7F0E}
\definecolor{tabgreen}{HTML}{2CA02C}
\definecolor{tabred}{HTML}{D62728}

\newcommand{\linesample}[2][black]{%
    \tikz[baseline=-0.6ex]\draw[#1,#2,line width=1pt] (0,0) -- (0.3,0);%
}

\newcommand{\blocksample}[2][black]{%
    \tikz[baseline={([yshift=-0.5ex]current bounding box.center)}]%
    \tikz[baseline=-0.3ex]\fill[#1,opacity=#2] (0,0) rectangle (0.35,0.28);%
}

\newcommand{\starsample}[1][tabblue]{%
  \tikz[baseline={([yshift=-0.5ex]current bounding box.center)}]
    \node[star, star points=5, star point ratio=2.3,
          draw=black, fill=#1, line width=0.5pt,
          minimum size=8pt, inner sep=0pt] {};%
}

\newcommand{\circlesample}[1][black]{%
  \tikz[baseline={([yshift=-0.5ex]current bounding box.center)}]
    \node[circle, draw=#1, fill=none, line width=1.0pt,
          minimum size=4pt, inner sep=0pt] {};%
}

\newcommand{\filledcirclesample}[1][black]{%
  \tikz[baseline={([yshift=-0.5ex]current bounding box.center)}]
    \node[circle, draw=#1, fill=#1, line width=1.0pt,
          minimum size=4pt, inner sep=0pt] {};%
}

\newcommand{\trianglesample}[1][black]{%
  \tikz[baseline={([yshift=-0.5ex]current bounding box.center)}]
    \node[regular polygon, regular polygon sides=3, draw=#1, fill=none,
          line width=1.0pt, minimum size=6pt, inner sep=0pt] {};%
}

\newcommand{\arrowsample}[2][black]{%
  \tikz[baseline=-0.5ex]\draw[#1,-{Stealth[length=5pt]},line width=0.8pt]
    (#2+180:0.2) -- (#2:0.1);%
}

\begin{document}

\begin{frontmatter}

\title{Reduced graph-based representations of detonation-containing flows}

\author[1]{Sinaendhran Pujali Elilarasan\texorpdfstring{\corref{cor1}}} %
\cortext[cor1]{Corresponding author}
\ead{spujalie@nd.edu}

\affiliation[1]{organization={Department of Aerospace and Mechanical Engineering, University of Notre Dame},%
            city={Notre Dame},
            postcode={46556}, 
            state={IN},
            country={USA}}

\author[2]{Michael Ullman} %

\affiliation[2]{organization={Department of Aerospace Engineering, University of Michigan},%
            city={Ann Arbor},
            postcode={48109}, 
            state={MI},
            country={USA}}

\author[1]{Shivam Barwey} %

\begin{abstract}
A graph neural network (GNN)-based autoencoder is leveraged to produce reduced latent graph representations of detonation-containing flows. In a two-dimensional hydrogen-air channel configuration with a uniform fuel-air mixture, it was found that latent graphs with effective resolutions of twice the induction length are capable of reconstructing both the wavefront and the post-shock region of the cellular detonation wave structure. Extrapolation studies using stratified and three-dimensional detonation configurations revealed an intrinsic property of GNN-based reconstruction from reduced latent graphs -- a representational hierarchy, rooted in the local nature of message passing, that tends to prioritize features at the detonation front (i.e., the wavefront profile and triple points) over the downstream deflagrative regions. In particular, it was found that at a fixed ambient condition, the information contained in uniform two-dimensional detonations was sufficient to reconstruct the wavefront structure of both stratified and three-dimensional detonations. Overall, this work demonstrates the utility of GNNs for both reduction and information overlap extraction in detonation-containing flows, opening promising pathways for accelerated simulations of these flows.
\end{abstract}

\begin{keyword}

Graph neural networks \sep Detonation \sep Machine learning \sep Complexity reduction
\end{keyword}

\end{frontmatter}

\tableofcontents

\section{Introduction}

Developing numerical models for accelerated simulation of detonation-containing flows is critical in a number of propulsion, energy, and safety applications \cite{venkat_arfm_rde,gulen_pgc_2017,short2018_explosive_detonation}. The modeling task here boils down to constructing a different \textit{reduced representation} of the high-fidelity detonation flow field, such that this representation provides (a) time-step gains (through elimination of prohibitive spatiotemporal scales), and (b) preservation of the information content of the resolved detonation. The balance between these two components of course varies based on the assumptions and methods used, each of which may target different physics for reduction (the detonation wave front, chemical kinetics, turbulence, or some combination) \cite{powers2006review}. 

For example, kinetics-oriented approaches range from simplified fictitious-species-based models \cite{short_quirk_jfm}, to QoI-driven mechanism reduction \cite{shepherd_mechanism_detonation}, in-situ tabulation-based modeling \cite{WU2018_isat_chem_acceleration}, and computational singular perturbation \cite{mauro_detonation}. While spatial resolution of the full wave structure is still required \cite{powers2005_lengthscale} in such approaches, acceleration comes from mitigation of temporal resolution costs stemming from detailed descriptions of chemical kinetics. On the other hand, approaches eliminating spatial resolution restrictions have been proposed, with the fundamental objective being to avoid resolving the detonation wave structure altogether. Strategies to this end include level-set-based modeling \cite{aslam1999_ghost_fluid}, such as detonation shock dynamics models \cite{short2016steady}, and shock fitting approaches \cite{keskinoz2026_shock-fitting}, which consolidate the detonation flow-chemistry interactions into a separately modeled (or experimentally-specified) detonation propagation speed. Reduced-order models for heat transfer have also been developed \cite{braun2018numerical,roy2016development}, as have turbulent combustion manifold-based models that operate in the detonation regime \cite{boerchers2026thermodynamically}. While individually powerful, the critical limitation of the above purely physics-based modeling approaches is that the assumptions underlying their reduction are tied to restrictive canonical configurations that may not apply in more realistic, complex-physics settings \cite{raman2019_emerging_trends}.

To mitigate this issue, data-driven modeling and machine learning (ML) is now actively being explored for combustion \cite{ml_combustion}. The critical advantage is that reduced representations -- referred to as \textit{latent} representations -- can be constructed directly from an optimization process on available high-fidelity flow data \cite{raman2019_emerging_trends}, providing a general pathway for representation learning that scales with increases in available data \cite{subramanian2023_scaling}. Application of data-driven modeling approaches to detonation-containing flows constitutes a promising direction to this end. Recognizing this, on the chemistry side, latent representations combining clustering with artificial neural networks have been used to develop models in simplified RDE and channel detonation configurations \cite{barwey2021_chem_source_est}. When considering the full detonation flowfield, latent representations based on modal decompositions, such as proper orthogonal decomposition (POD) and dynamic mode decomposition (DMD), have also been explored, wherein reduction is achieved using a reduced set of data-derived basis functions. Specifically, to accommodate the highly transient shock-driven process of a detonation, adaptive formulations of POD \cite{chenghuang2025_rde_aiaa, chenghuang2026_rde_scitech} and embedded DMD \cite{kutz2021_rde} have been developed for reduced-order modeling of rotating detonation engines. Additionally, accelerated models using convolutional neural network (CNN)-based latent representations have been leveraged for simulating shock-to-detonation transition \cite{nguyen2023_sdt} and 1D unsteady channel detonations \cite{nair2025_latent_timescales}. Recently, correction-based machine learning approaches based on sparse identification of nonlinear dynamics \cite{bao2026_cheap2rich, brunton2016_sindy} and data assimilation \cite{romit_rde_2026} have also been used to model rotating detonation engine dynamics.

The above approaches provide pathways for data-driven modeling and representation learning for detonations. Outside of detonations, to overcome limitations with either purely localized model forms, or projections tied to structured / fixed-grid assumptions (e.g., POD, DMD, or CNN-based neural networks), there has been increased adoption of mesh-based data-driven modeling incorporating graph neural networks in CFD \cite{pfaff2021_meshgraphnets,karthik_gnn}. Here, latent representations are produced over an unstructured graph whose adjacency matrix is derived from the mesh on which the high-fidelity flow field resides. The GNN, through neural message passing layers \cite{gilmer2017nmp, battaglia2018}, models information exchange in node neighborhoods using a multi-step aggregation procedure consisting of node and edge feature updates, with the aggregation lengthscale specified by the size of the local neighborhood. Critically, the GNN comes with an inductive bias: the learned aggregations are (a) neighborhood-local, and (b) explicit functions of the graph connectivity, which is provided as an input at the inference stage. This allows GNNs trained on one connectivity (mesh) to be applied to different connectivities (meshes) without retraining. While these qualities have been demonstrated in the broader fluid dynamics and physical science fields for mesh-based autoencoding \cite{barwey2023_gnnae,immordino2025_gcn}, autoregressive surrogate modeling \cite{jaiman_hypergraph,jaiman2025_gnn_surrogate,deshpande2024magnet}, closure modeling \cite{kim2026_gnn_closure_modelling,jianxun_gnn_solver}, super resolution \cite{barwey2025_super-resolution_cmame}, sensor reconstruction \cite{Li2024_flowfield_reconstruction}, and generative modeling \cite{lino2025_learningdistributionscomplexfluid} tasks, the application of GNNs towards modeling detonations, and the degree to which GNN-derived latent representations can generalize for detonation-containing flows, is yet to be explored. A detailed understanding of graph-based latent representations produced by GNN architectures can lead to improved pathways for accelerated modeling of these flows, and is the scope of the current study. 

The objective of the present work is to leverage a graph neural network (GNN)-based approach for the reduction of high-fidelity cellular detonations. To this end, a GNN autoencoder architecture is used. Reduced representations in this context correspond to latent graphs, whose nodes constitute what is effectively a coarse grid (i.e., the reduction here is resolution-based). The edges in the latent graph allow for specification of a characteristic length scale that can be related to the induction length of the detonation -- a critical length scale characterizing the chemical kinetics and grid resolution requirements in high-fidelity simulations. As such, this method provides a way to connect the expressive capability of the latent representation to an interpretable detonation quantity of interest. 

The present work sits alongside recent work in graph-based analysis and modeling for detonations: Ref.~\cite{vansh_lattice} leverages a graph representation to facilitate analysis of three-dimensional cellular detonation lattices, and Ref.~\cite{pinaki_srgt} explores mesh-based graphs for super-resolution of detonations using transformer architectures. The present work is distinct from, and adds to, this emerging body of work through its emphasis on a GNN autoencoding framework. Specifically, the primary goal is to assess the degree to which a reduced, latent graph representation can represent a high-fidelity detonation, particularly when (a) the latent graph length scale is several factors larger than the characteristic induction length of the detonation, and (b) latent representations are used in geometry- and physics-extrapolative contexts. This analysis provides the groundwork for future efforts as they relate to reduced-order modeling of detonations using graph-based latent representations. The specific contributions of this work are as follows: 
\begin{itemize}
    \item \textbf{Graph-based reduction of detonations:} A neural message passing (NMP)-based graph neural network autoencoder architecture is leveraged for generating latent graph-based representations of hydrogen-air cellular detonations in a channel configuration. A detailed analysis of the effect of the latent graph length scale on reconstruction fidelity is conducted and connected to the induction length, establishing the degree to which the fine-scale detonation wave structure can be recovered from a latent graph representation.
    \item \textbf{Stratified detonation extrapolation:} A GNN trained on uniform channel detonations is extrapolated to detonations propagating through stratified fuel-air mixtures. This comparison shows how GNNs can be used to isolate the information content shared between uniform and stratified detonation wave structures, while also quantifying the latent graph generalization to an out-of-distribution condition characteristic of more practical detonation engines.
    \item \textbf{Dimensional transferability of latent graphs:} The capacity of latent graph representations to transfer across spatial dimensions is examined by using a model trained on two-dimensional data to reconstruct three-dimensional detonations, both with and without transfer learning. This exposes the extent to which two-dimensional GNN-based latent representations of detonations can leverage self-similarity in the cellular structure to recover the complete 3D detonation.
\end{itemize}

The paper proceeds as follows. Section~\ref{sec:simulations_data} describes the simulation configurations and dataset. The architecture of the GNN model is described in Sec.~\ref{sec:methods}, followed by a discussion on the results and findings in Sec.~\ref{sec:results}. Concluding remarks are provided in Sec.~\ref{sec:conclusion}.

\section{Simulations and Data}
\label{sec:simulations_data}

\subsection{Numerical Configuration}
\label{sec:simulations}

The training and evaluation data used in this study comes from high-fidelity simulations of unsteady hydrogen-air channel detonations in two and three spatial dimensions. The governing equations for this configuration are the gas-phase compressible reacting Navier Stokes equations, which includes transport equations for mass, momentum, energy and species concentrations. The mass transport equation is given by
\begin{equation}
    \frac{\partial \rho}{\partial t} + \frac{\partial \rho u_i}{\partial x_i} = 0,
    \label{eq:mass_transport}
\end{equation}
where $\rho$ is the density of the fluid, and $u_i$ and $x_i$ are the component of velocity and the spatial location in the $i$th direction. The momentum transport equation is given by
\begin{equation}
    \frac{\partial \rho u_i}{\partial t} + \frac{\partial \rho u_j u_i }{\partial x_j} = -\frac{\partial p}{\partial x_i} + \frac{\partial \tau_{ij}}{\partial x_j}, 
    \label{eq:momentum_transport}
\end{equation}
where the viscous stress tensor $\tau_{ij}$ is 
\begin{equation}
    \tau_{ij} = -\frac{2}{3}\mu\frac{\partial u_k}{\partial x_k} \delta_{ij} + \mu \left( \frac{\partial u_j}{\partial x_i} + \frac{\partial u_i}{\partial x_j} \right),
    \label{eq:viscous_stress_tensor}
\end{equation}
 and the fluid viscosity, $\mu$, is computed based on the temperature of the fluid. The energy transport equation is given by
 \begin{equation}
     \frac{\partial \rho e}{\partial t} + \frac{\partial u_j \rho h}{\partial x_j} = \frac{\partial}{\partial x_j} \alpha \frac{\partial T}{\partial x_j} + \frac{\partial \tau_{ij} u_i}{\partial x_j} + \sum_{k=1}^{N_s} \left( h_k \frac{\partial}{\partial x_j} \left( \rho D \frac{\partial Y_k}{\partial x_j}\right) \right),
     \label{eq:energy_transport}
 \end{equation}
where $e$ is the total chemical energy, and $h = e + p/\rho$ is the corresponding total enthalpy. Here, $T$ is the temperature, $\alpha$ is the thermal conductivity of the mixture, $N_s$ is the number of chemical species, $Y_k$ and $h_k$ are the mass fraction and specific enthalpy of the $k$-th species, respectively, and $D$ is the spatially-varying species diffusivity (assumed to be equal for all species). The species transport equation is
\begin{equation}
    \frac{\partial \rho Y_k}{\partial t} + \frac{\partial u_i \rho Y_k }{\partial x_i} = \frac{\partial}{\partial x_i} \rho D \frac{\partial Y_k}{\partial x_i} + \Omega_k, \quad k = 1,..., N_s,
    \label{eq:species_mass_fraction_transport}
\end{equation}
where $\Omega_k$ is the chemical source term (species net production rate) for the $k$-th species.

The above equations, which are closed by the ideal gas equation of state, were solved using an AMReX-based compressible reacting flow solver~\cite{shivank_hmm} leveraging a detailed chemical kinetics model for the chemical source terms. The solver is an adaptive mesh refinement (AMR) and embedded boundary-based adaptation of UMReactingFlow \cite{bielawski2023_gpu_solver}, and has been used in a wide variety of detonation-related studies \cite{ullman2025_cnf_strat, barwey2025_chemical_timescales, SHARMA2024_JISC}. It is fully GPU-offloaded, and leverages a 2nd-order operator-split formulation with MUSCL-based reconstruction scheme for the advective fluxes, a central scheme for the diffusive fluxes, and an adaptive explicit method for chemical time integration. In this work, $H_2 - \text{air}$ detonations are considered using the mechanism of Mueller et al.\ \cite{muller1999} for chemical kinetics treatment. This mechanism consists of 9 species (N$_2$, O$_2$, H$_2$, H$_2$O, H, O, OH, HO$_2$, and H$_2$O$_2$) and 21 reactions, and has been used extensively for the analysis and simulation of detonation-containing flows \cite{prakash2020_lrde, ullman2024_detonation_strutures}.

\subsection{Description of Data}
\label{sec:dataset}
As mentioned above, both 2D and 3D channel configurations are used to produce data: the 2D data comes from the previous simulations of Ref.~\cite{ullman2025_cnf_strat}, which includes uniform and stratified detonations, while the 3D data are obtained from new simulations performed by the current authors. As such, the dataset is organized into the following three categories based on the configuration: \textbf{(1) 2D Uniform}: detonations propagating through a uniform $H_2$-air mixture in a 2D channel; \textbf{(2) 2D Stratified:} same as (1), but through a \textit{stratified} fuel-air mixture in 2D; and \textbf{(3) 3D Uniform:} same as (1), but through a uniform H2-air mixture in 3D. A more detailed description of each of the data categories is provided below.

\underline{\textbf{2D Uniform:}} The 2D Uniform dataset consists of snapshots sourced from Case 1 of Ref.~\cite{ullman2025_cnf_strat}, which correspond to unsteady $H_2 - \text{air}$ detonations propagating through a two-dimensional channel with uniform, stoichiometric fuel-air mixture at ambient conditions of 0.5~atm and 297~K. Snapshots are extracted in the regime of sustained cellular detonation propagation, and each snapshot is described by a 6-level AMR grid hierarchy. The base grid (Level 0) contains $1536\times256$ cells, corresponding to a grid resolution of 146.5 $\mu$m in $x$ and 156.3 $\mu$m in $y$. At each higher level, the grid size is refined by a factor of 2 for tagged cells at the previous level, resulting in 4.58 $\mu$m and 4.88 $\mu$m resolution at the highest level (Level 5). In the present study, the raw 6-level snapshots are preprocessed to produce the snapshots used for training and testing. First, from the full grid hierarchy, the data at refinement Level 1 (resolution of 73.25 $\mu$m and 78.15 $\mu$m) is extracted. Full level 1 snapshots are then partitioned into 10 sub-domains along the $y$-axis, each of which produces 10 cropped snapshots of size 11.65 mm in $x$, and 4 mm in $y$ used for training and testing.

\underline{\textbf{2D Stratified:}} The 2D Stratified dataset consists of snapshots sourced from Case 3 of Ref.~\cite{ullman2025_cnf_strat}, which correspond to unsteady $H_2-\text{air}$ detonations propagating through the same channel configuration described above, but through a \textit{stratified} fuel-air mixture rather than a uniform one. The stratification intensity is prescribed using a digital filtering approach~\cite{KLEIN2003_digital_filter}, resulting in an equivalence ratio for the stratified mixture ranging from $\phi = 0.6$ to $1.4$, with a mean of $1$. The computational grid matches that used for the 2D Uniform dataset. As with 2D Uniform, from the full grid hierarchy, only data from the Level 1 grid is used for the present study, and the same cropping procedure is used, producing subdomains of sizes 21.02 mm in $x$, and 4 mm in $y$.

\underline{\textbf{3D Uniform:}} The 3D Uniform dataset is the three-dimensional counterpart to 2D Uniform, and is generated by running new simulations using the same solver, with an added depth dimension of 7.6~mm along the $z$-axis. The base grid along the $z$-axis contains $32$ cells, producing a resolution of 218.75 $\mu$m in $z$, with the resolutions in $x$ and y remaining the same as 2D Uniform/Stratified. As with the 2D cases, the 3D cases were run with a total of 6 refinement levels, using an AMR tagging criteria based on the heat release rate (HRR) and the pressure gradient (cells with HRR greater than 200 GW/m$^3$ and pressure gradient greater than 2 GPa/m were tagged for refinement). This tagging criterion resulted in roughly 16 billion cells across all 6 refinement levels of the computational grid, compared to roughly 412 billion cells without AMR to achieve the same resolution as in the highest level. The simulations were run using 512 nodes (each with 12 effective GPUs) of the Aurora supercomputer at the Argonne Leadership Computing Facility (ALCF) \cite{aurora_alcf}. All 3D snapshots were extracted after stable detonation propagation at the Chapman-Jouguet (CJ) wavespeed. Analogous to the 2D cases, the Level 1 data is extracted from each 6-level 3D snapshot, and is then partitioned into 10 slices along the $y$-axis and 2 slices along the $z$-axis, producing 20 cropped, three-dimensional snapshots from each full snapshot, each of sizes 18.68 mm in $x$, 4 mm in $y$, and 3.8 mm in $z$. 

\begin{figure*}[ht]
    \centering
    \setlength{\abovecaptionskip}{-10pt}
    \includegraphics[width=1.0\linewidth]{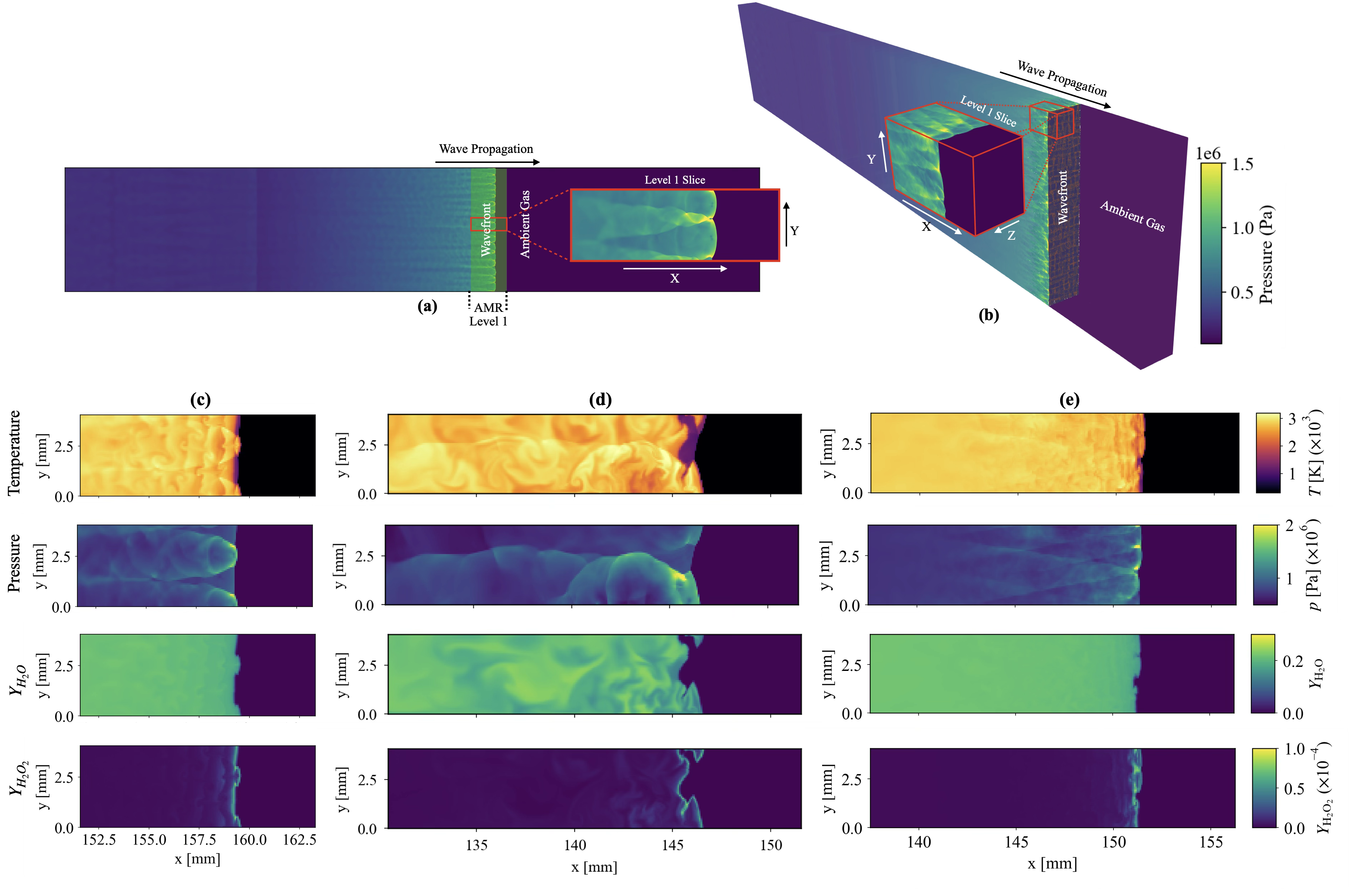}
    \caption{\textbf{(a)} A time snapshot of the pressure field from the 2D-Uniform dataset, showing the cropped Level 1 subdomain. \textbf{(b)} Same as (a), but for the 3D-Uniform dataset. \textbf{(c)} Cropped Level 1 temperature, pressure, $Y_{H_2O}$, and $Y_{H_2O_2}$ fields from the 2D-Uniform dataset. \textbf{(d)} Cropped fields from the 2D-Stratified dataset. \textbf{(e)} Fields from an x-y slice taken at the mid-plane of a cropped Level 1 data from the 3D-Uniform dataset.}
    \label{fig:snapshot_comparison}
\end{figure*}

The full collection of training, validation, and testing data snapshots for each case is provided in Table~\ref{tab:simulation_cases}. Each snapshot stores the full conserved state vector on each cell (mass, momentum, energy, and species concentrations). A comparison is shown in Fig.~\ref{fig:snapshot_comparison}, which illustrates the extraction of the cropped subdomains (final snapshots) from Level 1-tagged data. As seen in the figure, the domains (and number of grid points) of the final cropped snapshots \textit{vary} across the three simulation configurations because they were sourced from Level 1-tagged data -- the boundaries of which need not be the same across the configurations (owing to the differences in dynamics). For example, the $x$-domain extents for the stratified cropped snapshots are necessarily larger than uniform counterparts due to increased complexity behind the detonation front, and the 3D cropped snapshots are naturally different due to the presence of the third dimension. As will be shown further below, the graph neural network formulation can handle such domain and geometric variations. 

Given this cropping procedure, Fig.~\ref{fig:snapshot_comparison}(c-e), which shows flow visualizations for all three cases, illustrates how the final cropped snapshots are sufficiently large enough to retain all complexity of the detonation wavefront structure. Immediate inspection of pressure fields shows presence of triple point interactions of varying complexity -- due to the effects of fuel stratification, snapshots in the 2D Stratified cases (Fig.~\ref{fig:snapshot_comparison}(d)) are characterized by a more complex and corrugated wavefront structure than 2D Uniform (Fig.~\ref{fig:snapshot_comparison}(c)) and 3D Uniform (Fig.~\ref{fig:snapshot_comparison}(e)) counterparts. The $Y_{H_2O}$ plots show pockets of low and non-uniform concentrations of H$_2$O present in stratified detonations, indicative of a more complex wave structure. Figure~\ref{fig:snapshot_comparison} also highlights the key difference in the 3D setting -- namely, an extended reaction zone (as seen in the $Y_{H_2O_2}$ slice), which stems from transverse wave interactions along the $z$-axis.    

\begin{table*}[!ht]
    \centering
    \caption{Number of snapshots used for training, validation, and testing in all three cases (2D-Uniform, 2D-Stratified, and 3D-Uniform). All snapshots are cropped Level 1 subdomains extracted from the full AMR snapshots, as described in Sec.~\ref{sec:dataset} and shown in Fig.~\ref{fig:snapshot_comparison}.}
    \label{tab:simulation_cases}
    
    \begin{tabular}{{>{\centering\arraybackslash}p{2.5cm}>{\centering\arraybackslash}p{2.5cm}>{\centering\arraybackslash}p{2cm}>{\centering\arraybackslash}p{2cm}>{\centering\arraybackslash}p{2cm}}}
        \hline
        Dataset & Total snapshots (cropped) & Training snapshots & Validation snapshots & Testing snapshots \\
        \hline
        2D-Uniform      & 3960  & 3430 & 380 & 150\\
        2D-Stratified  & 3320 & 2853 & 317 & 150  \\
        3D-Uniform     & 61 & 54 & 6 & 1  \\
        \hline
    \end{tabular}
\end{table*}

\section{Methodology}
\label{sec:methods} 

With the description of data provided above, the graph-based reduction approach is now described. This is decomposed into two phases: (1) construction of graphs from detonation flowfields (Sec.~\ref{sec:graph_representation}), and (2) the graph neural network autoencoder acting on these graphs (Sec.~\ref{sec:gnn_architecture}). The former creates a graph representation of a given high-fidelity detonation snapshot; the latter is the model that learns a latent, reduced representation of this graph.

\subsection{Graph Representation of Flow}
\label{sec:graph_representation}

\begin{figure}
    \centering
    \setlength{\abovecaptionskip}{-10pt}
    \includegraphics[width=1.0\linewidth]{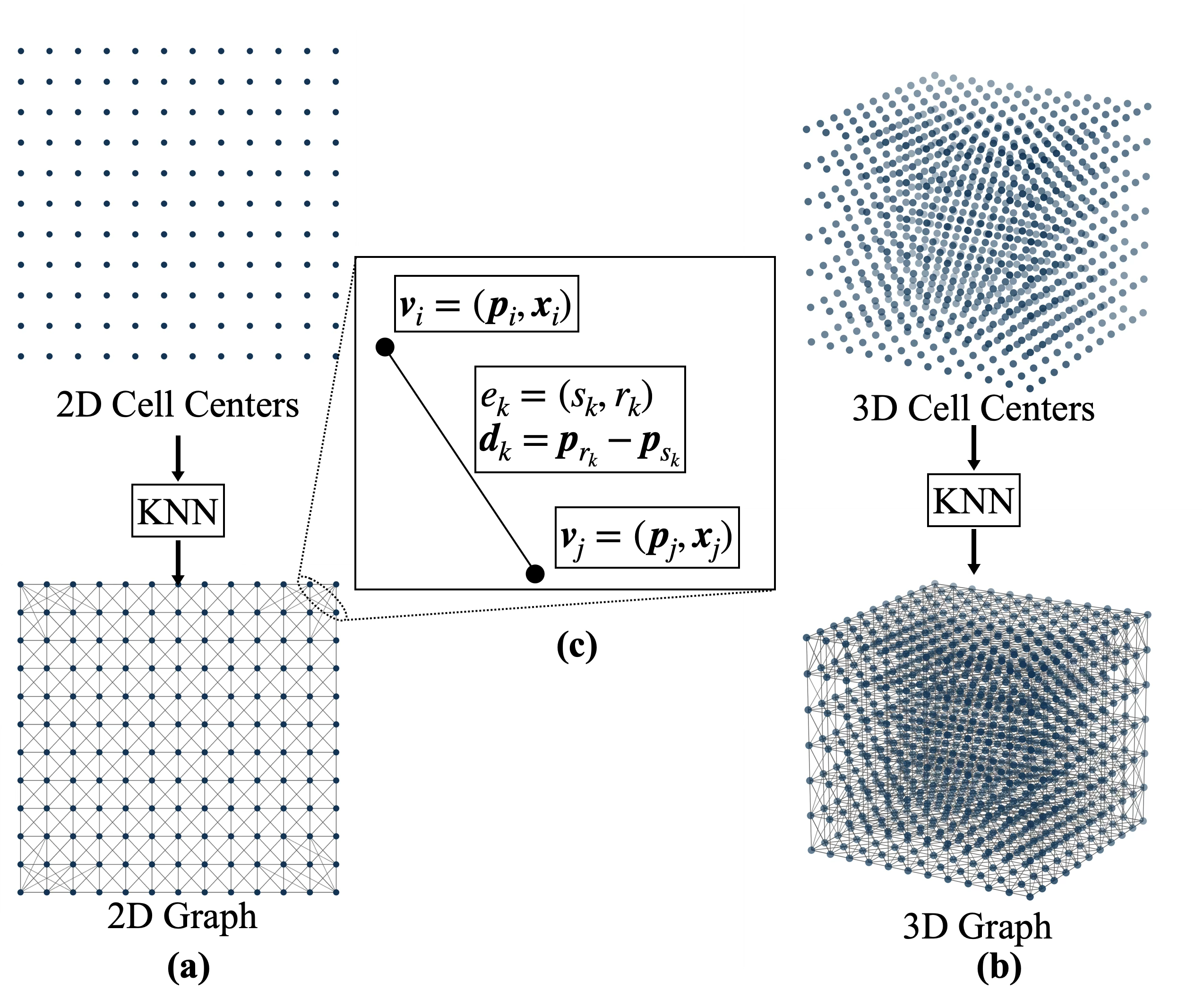}
    \caption{A schematic of graph construction from cell centers using the $k$-nearest neighbor (KNN) approach. \textbf{(a)} 2D graph construction. \textbf{(b)} 3D graph construction. \textbf{(c)} A pair of nodes and the edge connecting them.}
    \label{fig:2d_graph_generation}
\end{figure}

Any graph can be defined as $\mathcal{G} = (V, E)$, where $V = \{v_1, \ldots, v_{|V|}\}$ is the set of nodes and $E = \{e_1, \ldots, e_{|E|}\}$ is the set of edges. Specification of the attributes (features) that characterize the nodes, as well as the algorithm to produce the set of edges from these attributes, constitutes a complete description of the graph. For simplicity, the discussion below operates in the context of converting a single detonation snapshot into a graph (each snapshot has its own graph). A schematic of this process is provided in Fig.~\ref{fig:2d_graph_generation}.

\underline{\textbf{Node initialization:}}
The graph generation procedure follows the mesh-based strategy, wherein graph nodes coincide with the physical space locations of mesh/grid points of a snapshot \cite{pfaff2021_meshgraphnets,barwey2023_gnnae}. Each node $v_i$ is defined by two vectors, such that $v_i = ({\bm p}_i, {\bm x}_i)$. The quantity ${\bm p}_i \in \mathbb{R}^{N_d}$ is the node \textit{position} vector, storing the $N_d$ Cartesian coordinates of the underlying grid point -- $N_d=3$ here, with the last entry set to $0$ for two-dimensional cases. The quantity ${\bm x}_i \in \mathbb{R}^{N_f}$ is the node \textit{feature} vector, storing $N_f$ non-positional node attributes, which here are the variables defining the local detonation flow and thermochemical state. These features coincide with the conserved variables: ${\bm x}_i = [\rho, \, \rho U_x, \, \rho U_y, \, \rho U_z, \, \rho E, \, \rho Y_k]$, for $k = 1, \ldots, N_s$, producing $N_f=14$ node features.

\underline{\textbf{Edge initialization:}} The $k$-th edge is defined as the tuple $e_k = (s_k, r_k)$, where $s_k$ and $r_k$ are the indices of the sender and receiver nodes respectively. The sender and receiver indices come from the $|V| \times |V|$ adjacency matrix, represented in practice using a sparse format as ${\bm A} \in \mathbb{Z}^{2 \times |E|}$, with the first row storing the sender indices and the second row storing the receiver indices for each of the edges. An edge construction step is required to construct ${\bm A}$. Edge construction provides an inductive bias to the graph neural network: an edge connecting two nodes encodes a similarity between them based on a distance measure, and the GNN layers propagate information along the pathways provided by these edges. Here, edges are constructed using a $k$-nearest neighbor (KNN) algorithm based on Euclidean distances in physical space, where an edge is created for every K-closest nodes to a given source node based on physical space proximity ($K=8$ is used here). Then, using the connectivity information in $E$, edge attributes are initialized using relative node positions as
\begin{equation}
    {\bm d}_k = {\bm p}_{r_k} - {\bm p}_{s_k} \in \mathbb{R}^{N_d}, \quad k = 1,\ldots,|E|.
\end{equation}

Any collection of node and edge attribute vectors can be consolidated into matrices for ease of representation, where the rows span the number of nodes $|V|$ and edges $|E|$, respectively. For example, the node feature matrix is ${\bm X} \in \mathbb{R}^{|V| \times N_f}$, with each row containing the corresponding feature vector ${\bm x}_i$ for the corresponding node -- analogous edge feature matrices can also be constructed. 

Here, it is emphasized that the graphs are constructed from cropped level-1 data and that, during the simulations, the computational mesh evolves over time, changing the number of level-1 cells at each coarse time step, resulting in slices of varying numbers of cells. The graph generation procedure results in roughly 8,000 nodes and 64,000 edges for each 2D-uniform graph, 13,000 nodes and 100,000 edges for each 2D-stratified graph, and 200,000 nodes and 1,600,000 edges for each 3D-uniform graph.

\subsection{Graph-based Reduction Approach}
\label{sec:gnn}

Section~\ref{sec:graph_representation} described the graph construction process for the input high-fidelity detonation snapshot. In the discussion hereafter, this input graph will be referred to as ${\cal G}_f = (V_f, E_f)$ -- the \textit{fine} graph. Alongside ${\cal G}_f$, there exists a reduced representation of this fine graph, termed the \textit{latent} graph ${\cal G}_l = (V_l, E_l)$, defined by a different, reduced set of nodes and edges than those in ${\cal G}_f$, such that $|V_l| < |V_f|$.

\subsubsection{Obtaining the Latent Graph}
\label{sec:latent_graph}
\begin{figure}[ht]
    \centering
    \setlength{\abovecaptionskip}{-10pt}
    \includegraphics[width=1.0\linewidth]{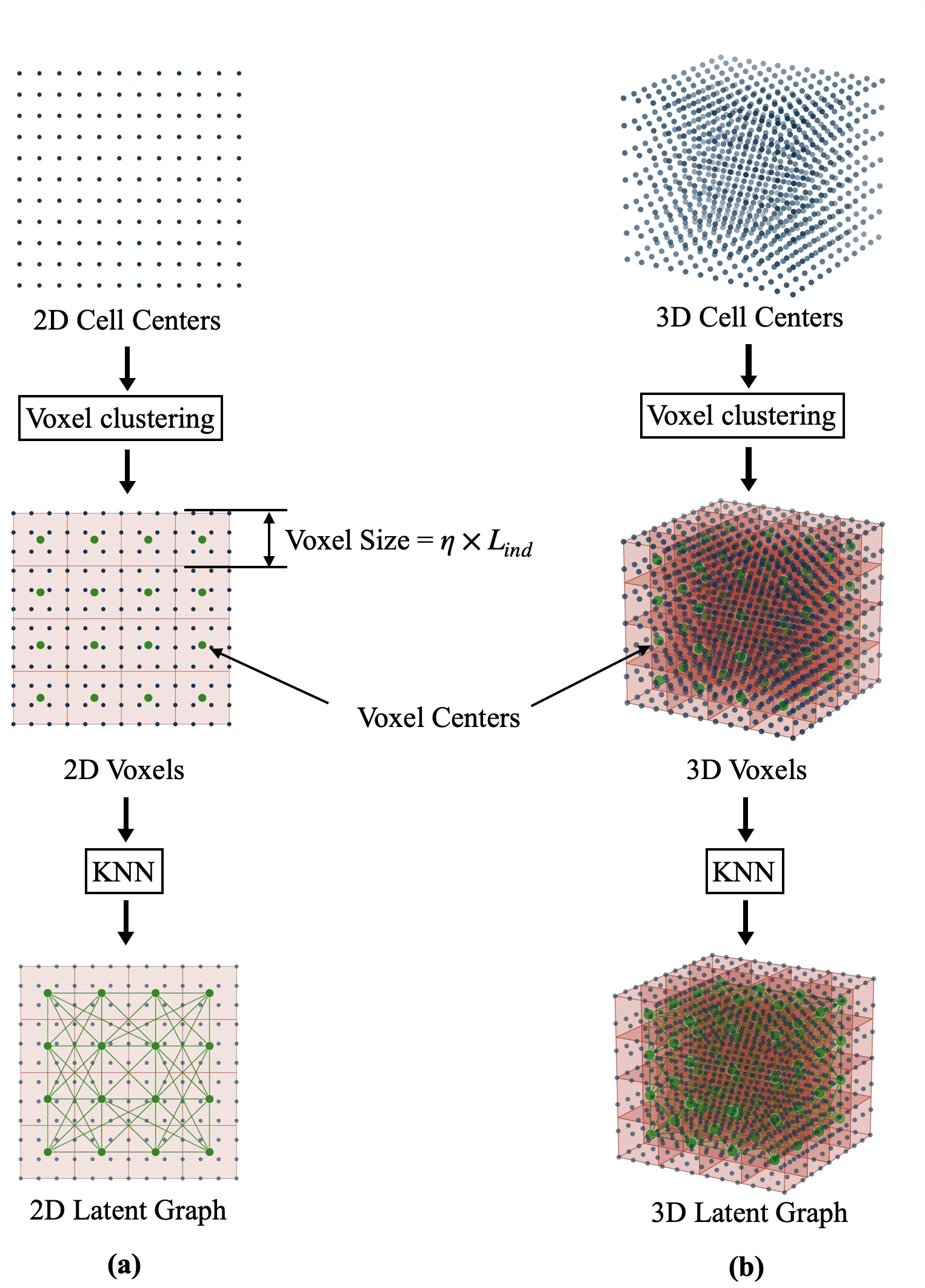}
    \caption{A schematic of latent graph construction using the voxel clustering approach. \textbf{(a)} 2D latent graph construction from 2D cell centers, showing fine graph nodes (blue), voxels (red boxes), and latent graph nodes at the voxel centers (green points) connected by latent graph edges (green lines). \textbf{(b)} Same as (a), but for 3D latent graph construction from 3D cell centers.}
    \label{fig:latent_graph}
\end{figure}

A voxel clustering approach applied to the fine-graph node position matrix, ${\bm P}_f$, is used to obtain the nodes $V_l$ and edges $E_l$ of the latent graph \cite{lino_multiscale_gnn, barwey2023_gnnae}. The physical domain is divided into squares (for 2D graphs) or cubes (for 3D graphs), termed voxels, of user-specified size $\eta \times L_{ind}$, where $L_{ind}$ is the characteristic induction length of the detonation (derived from Zel'dovich-von Neumann-Döering (ZND) model in SDToolbox \cite{edl_sdtoolbox_2023}) and $\eta$ is the ratio of voxel size to the induction length. This step assigns each fine-graph node an integer voxel label, and the center of each occupied voxel defines the location of a corresponding latent node. \textbf{As such, the latent graph nodes coincide with a coarse mesh, whose resolution is set by the nondimensional parameter $\eta$, providing a direct way to relate the latent graph encoding ability to the detonation induction length $\boldsymbol{L}_{ind}$}. A schematic is provided in Fig.~\ref{fig:latent_graph}. 

The spatial position of the $j$-th latent graph node, ${\bm p}_{l,j} \in \mathbb{R}^{N_d}$, is obtained via a graph pooling operation that uses the voxel labels to compute the centroid of the fine-graph node positions assigned to each voxel,
\begin{equation}
    \label{eq:voxel_pooling}
    {\bm p}_{l,j} = \frac{1}{|\mathcal{C}_j|} \sum_{i \in \mathcal{C}_j} {\bm p}_{f,i},
\end{equation}
where $\mathcal{C}_j$ is the set of fine-graph node indices sharing the $j$-th voxel label, $|\mathcal{C}_j|$ is the number of such nodes, and ${\bm p}_{f,i}$ is the position vector of node $i$ on the fine graph ${\cal G}_f$. The latent graph edges, $E_l$ are constructed using a similar KNN algorithm, with the edge attributes initialized using the pooled node attributes as 

\begin{equation}
    \label{eq:latent_edge_init}
   {\bm d}_{l,k} = \tilde{\bm x}_{l,r_k} - \tilde{\bm x}_{l,s_k} \in \mathbb{R}^{N_h}, \quad k = 1,...,\left| E_l \right|, 
\end{equation}
where $\tilde{\bm x}_{l,r_k}$ and $\tilde{\bm x}_{l,s_k}$ are the corresponding rows in the embedded latent node feature matrix, $\tilde{\bm X}$.

\subsubsection{GNN Autoencoder}
\label{sec:gnn_architecture}

\begin{figure*}[ht]
    \centering
    \setlength{\abovecaptionskip}{4pt}
    \includegraphics[width=0.8\linewidth]{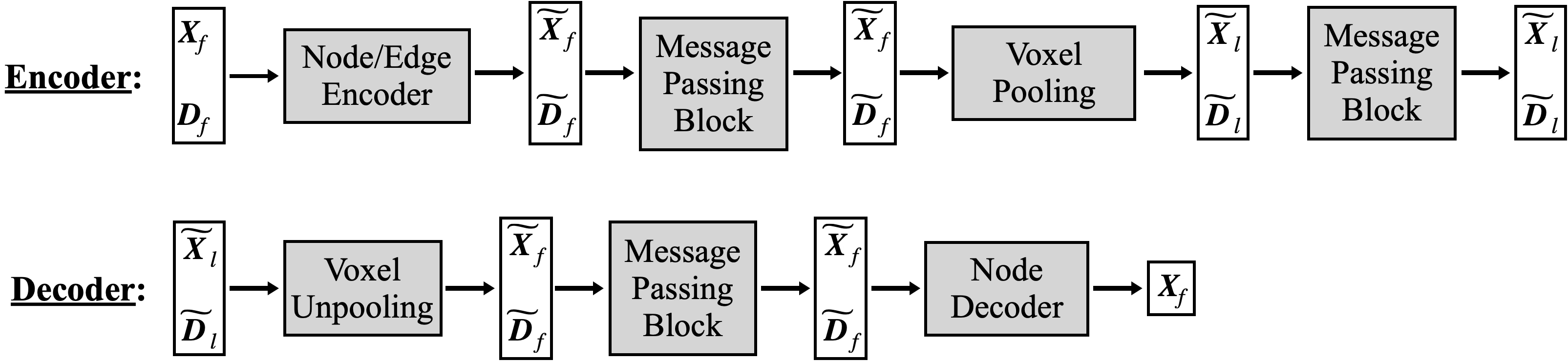}
    \caption{Graph neural network (GNN)-based autoencoder architecture. Each component is described in Sec.~\ref{sec:gnn_architecture}.}
    \label{fig:gnn_architecture}
\end{figure*}

A schematic of the GNN architecture is presented in Fig.~\ref{fig:gnn_architecture}. It follows an autoencoding approach, consisting of an encoder and a decoder. Each component is described in succession below.

\underline{\textbf{Encoder:}} The encoder algorithm is provided in Alg.~\ref{alg:encoder}. The goal of the encoder is to transform the input representation of the flow field (the node feature matrix on ${\cal G}_f$) into a latent representation (a node feature matrix on ${\cal G}_l$).
First, the input node and edge feature vectors are encoded into a hidden dimension using independently parameterized multi-layer perceptrons (MLPs) acting row-wise over the respective matrices (Alg.~\ref{alg:encoder}, Lines 1--2). The embedded matrices are $\widetilde{\bm X}_f \in \mathbb{R}^{|V_f| \times N_h}$ and $\widetilde{\bm D}_f \in \mathbb{R}^{|E_f| \times N_h}$, with the latent feature (or ``hidden'') dimension $N_h = 128$ unless otherwise specified.
Second, a message passing block (MPBlock) operating on the fine-graph connectivity ${\bm A}_f$ updates these encoded node and edge features through non-local neighborhood aggregations over ${\cal G}_f$ (Alg.~\ref{alg:encoder}, Line 3). Then, after initializing the latent node and edge features through the voxel pooling strategy (Alg.~\ref{alg:encoder}, Lines 4--5), a second message passing block generates the final encoder output (Alg.~\ref{alg:encoder}, Line 6). This second MPBlock operates on the \textit{latent} connectivity ${\bm A}_l$ instead of the fine connectivity ${\bm A}_f$, and is used to generate the final latent node features. The resulting updated $\widetilde{\bm X}_l$ is the final latent graph-based representation of the detonation.

\begin{algorithm}[ht]
\caption{Encoder}
\label{alg:encoder}
\begin{algorithmic}[1]
\Require Fine graph node features ${\bm X}_f$, fine graph edge features ${\bm D}_f$, fine graph connectivity ${\bm A}_f$, latent graph connectivity ${\bm A}_l$, voxel labels $\mathcal{C}$
\State \textbf{Node feature encode:} $\widetilde{\bm X}_f  \leftarrow \text{MLP}({\bm X}_f)$
\State \textbf{Edge feature encode:} $\widetilde{\bm D}_f  \leftarrow \text{MLP}({\bm D}_f)$
\State \textbf{Message passing:} $\widetilde{\bm X}_f, \widetilde{\bm D}_f \leftarrow \text{MPBlock}(\widetilde{\bm X}_f, \widetilde{\bm D}_f, {\bm A}_f)$
\State \textbf{Latent node init:} $\widetilde{\bm X}_l \leftarrow \text{VoxelPool}(\widetilde{\bm X}_f, \mathcal{C})$ (via Eq.~\ref{eq:voxel_pooling})
\State \textbf{Latent edge init:} $\widetilde{\bm D}_l \leftarrow \widetilde{\bm X}_{l,r_k} - \widetilde{\bm X}_{l,s_k} $ (via Eq.~\ref{eq:latent_edge_init})
\State \textbf{Message passing:} $\widetilde{\bm X}_l, \widetilde{\bm D}_l \leftarrow \text{MPBlock}(\widetilde{\bm X}_l, \widetilde{\bm D}_l, {\bm A}_l)$
\State \Return $\widetilde{\bm X}_l \in \mathbb{R}^{|V_c| \times N_h}$
\end{algorithmic}
\end{algorithm}

\begin{algorithm}[ht]
\caption{Decoder}
\label{alg:decoder}
\begin{algorithmic}[1]
\Require Latent node features $\widetilde{\bm X}_l$, latent edge features ${\bm D}_l$, latent graph connectivity ${\bm A}_l$, fine graph connectivity ${\bm A}_f$, fine graph node positions ${\bm P}_f$, coarse graph node positions ${\bm P}_l$
\State \textbf{Fine node init:} $\widetilde{\bm X}_f \leftarrow \text{KNNInterp}(\widetilde{\bm X}_l, {\bm P}_l, {\bm P}_f)$ (via Eq.~\ref{eq:knn_interpolation})
\State \textbf{Fine edge init:} $\widetilde{\bm D}_f \leftarrow \widetilde{\bm X}_{f,r_k} - \widetilde{\bm X}_{f,s_k}$ (via Eq.~\ref{eq:latent_edge_init})
\State \textbf{Message passing:} $\widetilde{\bm X}_f, \widetilde{\bm D}_f \leftarrow \text{MPBlock}(\widetilde{\bm X}_f, \widetilde{\bm D}_f, {\bm A}_f)$
\State \textbf{Node feature decode:} $\widetilde{\bm X}_f \leftarrow \text{MLP}(\widetilde{\bm X}_f)$
\State \Return $\widetilde{\bm X}_f \in \mathbb{R}^{|V_f| \times N_f}$
\end{algorithmic}
\end{algorithm}

\underline{\textbf{Decoder:}} The decoder algorithm is provided in Alg.~\ref{alg:decoder}. The goal of the decoder is to supply a reconstruction $\widetilde{\bm X}_f \approx {\bm X}_f$ using information contained only in the latent graph-based representation of the flow, $\widetilde{\bm X}_l$.
First, an unpooling operation interpolates these coarse graph features onto the fine graph (Alg.~\ref{alg:decoder}, Line 1). A KNN-interpolation approach is used, where the feature at fine-graph node $i$ is recovered from a distance-weighted average of the features at its $k$ closest nodes on the coarse graph,
\begin{equation}
    \label{eq:knn_interpolation}
    \tilde{\bm x}_{f,i} = \frac{\sum_{m=1}^k b_m \, \tilde{\bm x}_{l,m}}{\sum_{m=1}^k b_m}, \quad \text{where} \,\, b_m = \frac{1}{\lVert {\bm p}_{l,m} - {\bm p}_{f,i} \rVert^2_2},
\end{equation}
with ${\bm p}_{l,m}$, for $m \in \{1,\ldots,k\}$, denoting the positions of the $k$ closest coarse-graph nodes to the fine-graph node at ${\bm p}_{f,i}$ ($k=2$ is used here). Second, the fine graph edge attributes are reinitialized using the same procedure as in Eq.~\ref{eq:latent_edge_init}, with the interpolated, embedded fine graph node features, $\widetilde{\bm X}_f$ (Alg.~\ref{alg:decoder}, Line 2).
Third, a message passing block operating on the fine graph connectivity ${\bm A}_f$ propagates and mixes information among the fine nodes (Alg.~\ref{alg:decoder}, Lines 3). Fourth, a node feature decoder MLP maps the resulting fine-graph features from the latent space back to the physical feature space (Alg.~\ref{alg:decoder}, Lines 4--5), producing the reconstructed node feature matrix $\widetilde{\bm X}_f$.

\underline{\textbf{Message Passing Block:}}
Each message passing block (MPBlock) consists of a sequence of Neural Message Passing (NMP) layers \cite{gilmer2017nmp}, which are used extensively in the scientific machine learning community for mesh-based modeling purposes \cite{pfaff2021_meshgraphnets, barwey2025_super-resolution_cmame, gao2022_gnn_solver, gao_jaiman_2026_hgnn}. The NMP layer models information exchange through learnable, alternating node and edge feature updates; these updates are coupled through a neighborhood aggregation procedure that is conditioned on the graph connectivity. As such, the NMP layer is a fundamental model that allows one to impose a structure for information propagation through a graph. After the action of an NMP layer, the result is an updated set of node and edge features over the same fixed graph connectivity that is supplied as input. \textit{Critically, the trainable parameters of the NMP layers are not constrained to the connectivity}, which is what allows NMP-based architectures to extrapolate to different domains and, as will be shown here, spatial dimensions.

For a generic graph $\mathcal{G} = (V, E)$ with node feature vectors ${\bm x}_i$ and edge feature vectors ${\bm d}_k$ (connecting sender node $s_k$ to receiver node $r_k$), a set of $N_{mp}$ NMP layers forms a message-passing block. Each NMP layer performs the three successive operations listed below,
\begin{equation}
    \label{eq:nmp_edge_update}
    \text{Edge update:} \quad {\bm d}^{\,t}_{k} = {\bm d}^{\,t-1}_{k} + \text{MLP} \, ({\bm x}^{t-1}_{s_k}, \, {\bm x}^{t-1}_{r_k}, {\bm d}_k^{t-1}; \theta_e^t),
\end{equation}
\begin{equation}
    \label{eq:nmp_edge_agg}
    \text{Edge aggregation:} \quad {\bm a}^{\,t}_i = \sum_{k \,:\, r_k = i} {\bm d}^{\,t}_{k},
\end{equation}
\begin{equation}
    \label{eq:nmp_node_update}
    \text{Node update:} \quad {\bm x}^{t}_{i} = {\bm x}^{t-1}_{i} + \text{MLP} \, ({\bm x}^{t-1}_i, \, {\bm a}^{\,t}_i; \theta_n^t),
\end{equation}
where $\theta_e^t$ and $\theta_n^t$ are the parameters of the MLPs that update the edge and node features, respectively, at the $t$-th NMP layer. In the edge update step (Eq.~\ref{eq:nmp_edge_update}), the edge features ${\bm d}^{\,t}_{k}$ are updated by an MLP using the attributes of the sender and receiver nodes connected by that edge, from the previous layer. The edge aggregation step (Eq.~\ref{eq:nmp_edge_agg}), performed at every node $i$, sums the updated features of all edges in its neighborhood, producing the node-based representation ${\bm a}_i^t$. The NMP layer concludes with the node update step (Eq.~\ref{eq:nmp_node_update}), wherein each node is updated by an MLP using the aggregated edge attributes at that node alongside its previous attribute.

\subsection{Objective function}
The objective function is given by
\begin{equation}
    \label{eq:total_loss}
    \mathcal{L}_{total} = \mathcal{L}_{data} + \lambda\mathcal{L}_{elemental}
\end{equation}
where $\lambda$ is a weighting parameter, $\mathcal{L}_{data}$ is the standard reconstruction loss, and $\mathcal{L}_{elemental}$ is an elemental mass fraction penalty term \cite{WANG2025_elemental_loss, kumar_pal_2025_ChemNODE}. The data loss, computed as the Mean Squared Error (MSE) between the predicted and true node attributes, is given by
\begin{equation}
    \label{eq:data_loss}
    \mathcal{L}_{data} = \frac{1}{|V_f| \times N_f} \sum_{i=1}^{|V_f|} \sum_{j=1}^{N_f} ({\bm X}_{ij} - \widetilde{{\bm X}}_{ij})^2,
\end{equation}
and the elemental loss, computed as the MSE in the mass fractions of individual atoms at each node, is given by
\begin{equation}
    \label{eq:elemental_loss}
    \mathcal{L}_{elemental} = \frac{1}{|V_f| \times N_E} \sum_{i=1}^{|V_f|} \sum_{j=1}^{N_E} ({\bm Z}_{ij} - \widetilde{\bm {Z}}_{ij})^2,
\end{equation}
where the elemental mass fractions, ${\bm Z} \in \mathbb{R}^{N_E \times |V_f|}$, are computed as
\begin{equation}
    \label{eq:elemental_massfrac}
    {\bm Z}_{ik} = \sum_{j=1}^{N_S} {\bm A}_{ij} {\bm Y}_{jk}.
\end{equation}
Here, ${\bm A} \in \mathbb{R}^{N_E \times N_S}$ is the constant elemental mass fraction matrix, where $N_E$ is the number of individual elements, $N_S$ is the number of molecular species, and ${\bm Y} \in \mathbb{R}^{N_S \times |V_f|}$ is the matrix of species mass fractions. In both Eq.~\ref{eq:data_loss} and Eq.~\ref{eq:elemental_loss}, $\widetilde{(\cdot)}$ denotes the predicted (reconstructed) quantity, while $(\cdot)$ without the tilde denotes the true quantity. The factor $\lambda=100$ is used in Eq.~\ref{eq:total_loss}.

\section{Results}
\label{sec:results}

The models are implemented in PyTorch~\cite{paszke2019_pytorch} and PyTorch Geometric \cite{paszke2019_pytorch, Fey/etal/2025_pyG}. A cosine learning rate scheduler is used with a starting learning rate of $10^{-4}$ and an ending rate of $10^{-8}$. The AdamW optimizer \cite{adamw_optimizer} is used with a maximum gradient clipping of gradient value 1. All the models are trained for 70 epochs, with a batch size of 1 graph per batch, using 2 nodes (24 effective GPUs) of the Aurora high-performance computing cluster at the Argonne Leadership Computing Facility (ALCF) \cite{aurora_alcf}. Unless otherwise specified, the following model settings are used for the results below: the number of NMP layers per MP block is $N_{mp}=9$, the hidden feature dimension is ${N_h} = 128$, and all MLPs are constructed with 2 fully connected layers (excluding the input layer) using the ReLU activation function. This produces approximately 5 million trainable parameters in the GNN. 

The results proceed as follows. In Sec.~\ref{sec:lengthscale_analysis}, the GNN trained on 2D-Uniform data is evaluated in the same configuration in order to isolate the dependency of detonation wave structure reconstruction on the latent graph lengthscale. Then, in Sec.~\ref{sec:extrapolation}, an analysis of 2D-Uniform-trained models in physics extrapolation contexts (namely, on 2D-Stratified and 3D-Uniform data) is performed using the optimal-lengthscale GNN discussed in Sec.~\ref{sec:lengthscale_analysis}. %

In this section, to facilitate fine-scale wave structure analysis in detail, the models are evaluated using a representative test snapshot from each of the respective data categories. To supplement this, \ref{appendix:stat_analyses} provides global results of models trained on 2D-Uniform and 2D-Stratified data across all available test graphs (see Table~\ref{tab:simulation_cases}).

\subsection{Analysis of the Latent Graph Lengthscale}
\label{sec:lengthscale_analysis}

\begin{figure*}[ht]
    \centering
    \setlength{\abovecaptionskip}{5pt}
    \includegraphics[width=0.95\linewidth]{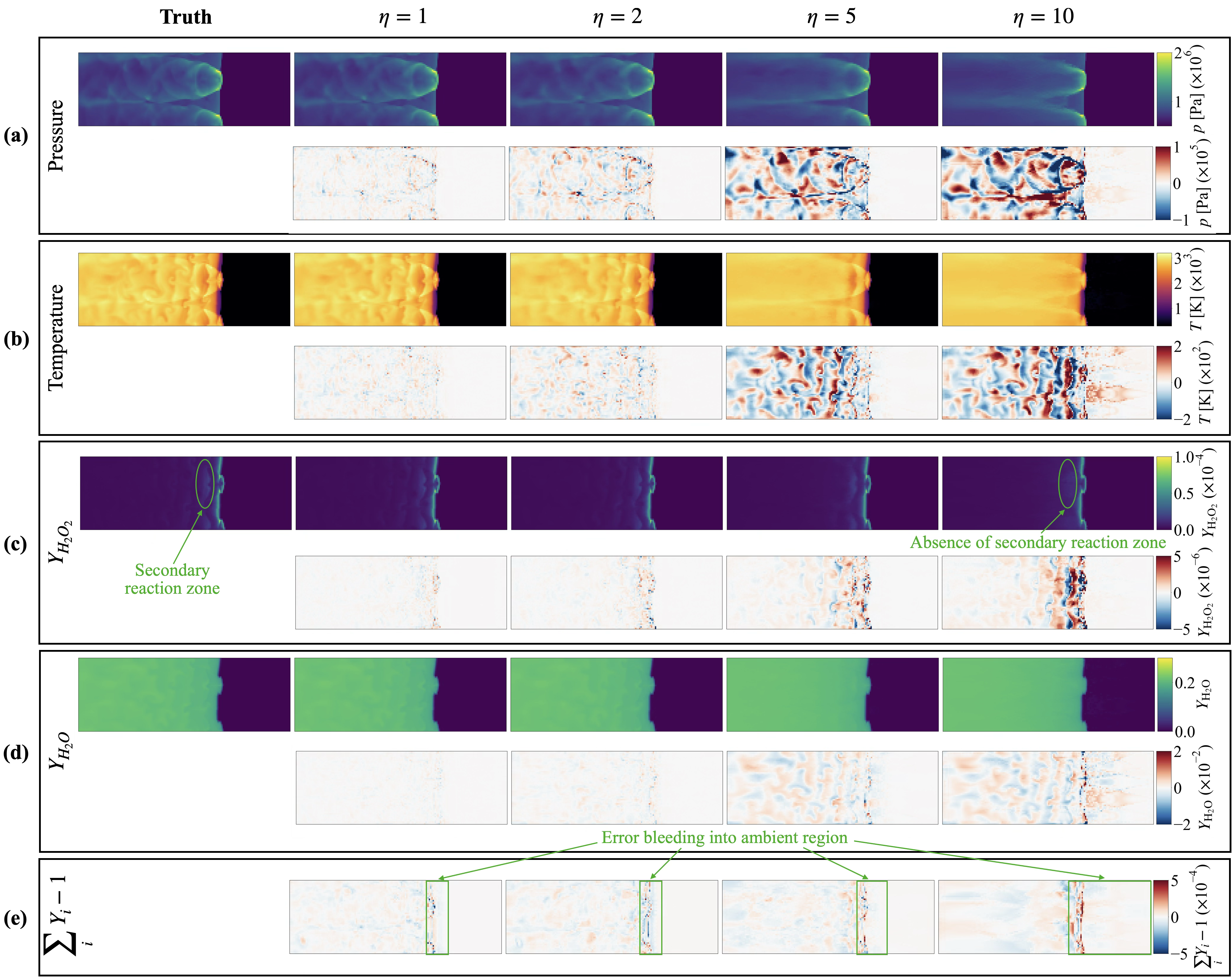}
    \caption{Reconstructions of a 2D-Uniform test-set snapshot for \textbf{(a)} pressure, \textbf{(b)} temperature, \textbf{(c)} $Y_{H_2O_2}$, and \textbf{(d)} $Y_{H_2O}$, with \textbf{(e)} showing species mass conservation error fields, $\sum_i Y_i - 1$. In each of (a)--(d), the top row shows the true field followed by predictions from models with latent graph length scales $\eta=1$, $2$, $5$, and $10$ (left to right), and the bottom row shows the corresponding error fields (difference between predicted and true fields).}
    \label{fig:field_predictions_over_eta}
\end{figure*}

Predicted pressure, temperature, H$_2$O$_2$ mass fraction, H$_2$O mass fraction, and mass conservation fields are presented in Fig.~\ref{fig:field_predictions_over_eta} for a series of latent graph length scales, $\eta \in \{1, 2, 5, 10\}$. On immediate inspection, it can be observed that as $\eta$ increases, the overall error in the decoded detonation increases, especially in the post-shock downstream structures. This highlights an expected trend, connecting the reconstruction capacity of the message-passing layers in the decoder to the latent graph resolution. For a fixed number of message-passing layers, at a coarser latent graph length scale, the finer-scale features of the detonation are more difficult for the decoder to recover.

While a general correlation between the latent graph length scale and the reconstruction capability exists, more detailed inspection of the trends in Fig.~\ref{fig:field_predictions_over_eta} reveals a more intricate trade-off. More specifically, a stark cutoff exists at $\eta = 2$: models trained using latent graph length scales at and below this number (amounting to effective resolutions up to $2\times$ the ZND induction zone length) successfully reproduce the downstream pressure and temperature fluctuations of the detonation, alongside the triple-point-driven structure at the leading shock wave. Assessment of the pressure and temperature fields alone thus indicates that a reduced latent graph operating at a resolution of twice the ZND induction length can carry sufficient information to allow the decoder to reconstruct the full detonation wave structure at the ambient conditions considered here. On the other hand, at $\eta = 5$ and $\eta = 10$, due to the increased coarsening, the downstream variation in the wave structure is completely killed; however, even for the $\eta = 10$ latent graph (an effective resolution of $10\times$ the ZND induction length), the overall structure of the leading shock, \textit{including the positions of the triple points and the transverse waves emanating from them}, is still recovered.

This alludes to a less obvious relationship between message-passing-based reconstruction and detonation physics: the reduction of latent graph resolution (i.e., movement from $\eta = 1$ to $\eta = 10$) manifests as a loss of post-shock flow information in order to pin the location of the triple points and transverse waves. In other words, this implies that the most salient features of the detonation wave, aside from the leading shock, are the triple points, which are indeed the driving features of a cellular detonation \cite{shepherd_detonation_in_gases_2009}.

Inspection of species fields provides additional insight into the reaction structure. In particular, Fig.~\ref{fig:field_predictions_over_eta}(c) and (d) show $Y_{H_2O_2}$ and $Y_{H_2O}$ reconstructions, respectively. All models consistently capture an intermediate species, $Y_{H_2O_2}$, in the primary reaction zone near the wavefront. However, models with higher latent graph lengthscales fail to capture this intermediate species in the secondary reaction zones, which are slightly behind the wavefront (as highlighted in Fig.~\ref{fig:field_predictions_over_eta}(c)). Especially for $\eta=10$, the secondary reaction zone is essentially killed. The error plots of $Y_{H_2O}$ reveal additional information about the role of these secondary reaction zones. As the latent graph length scale increases, the error becomes more pronounced in the post-shock region as corrugated regions, while reconstructing a more uniform $Y_{H_2O}$ field. This behavior can be attributed to the absence of secondary reaction zones, resulting in a more uniform H$_2$O field generated in the primary reaction zone, with no further H$_2$O production downstream.

Figure~\ref{fig:field_predictions_over_eta}(e) shows errors in the sum of mass fractions (i.e., conservation errors) at each graph node. A generic trend in the conservation plots is that as the coarsening level of the latent graph increases, the mass conservation error propagates more into the unreacted ambient region, which is highlighted in Fig.~\ref{fig:field_predictions_over_eta}(e). For models with $\eta \in \{1, 2\}$, the error is confined to the post-shock region and extends slightly into the ambient region near the wavefront, where there are more species present and the thermochemical gradients are higher. However, for models with higher $\eta$, the error propagates much farther into the unreacted region and is most prominent in the $\eta=10$ model. A similar trend appears in the other fields. This may be due to insufficient latent-graph resolution at the wavefront, leading to KNN interpolation errors in the decoder, which the message-passing layers later try to correct. In models with higher $\eta$, these interpolation errors are higher and not fully corrected by the message-passing layers. This is more pronounced when the nodes connecting the edges that cut the shock are farther apart, as at $\eta = 10$. For this configuration, $\eta=2$ appears to be a threshold for overall balance in capturing the secondary reaction zones and confining the errors to the post-shock region.

\begin{figure}[t]
    \centering
    \setlength{\abovecaptionskip}{-10pt}
    \includegraphics[width=1.0\linewidth]{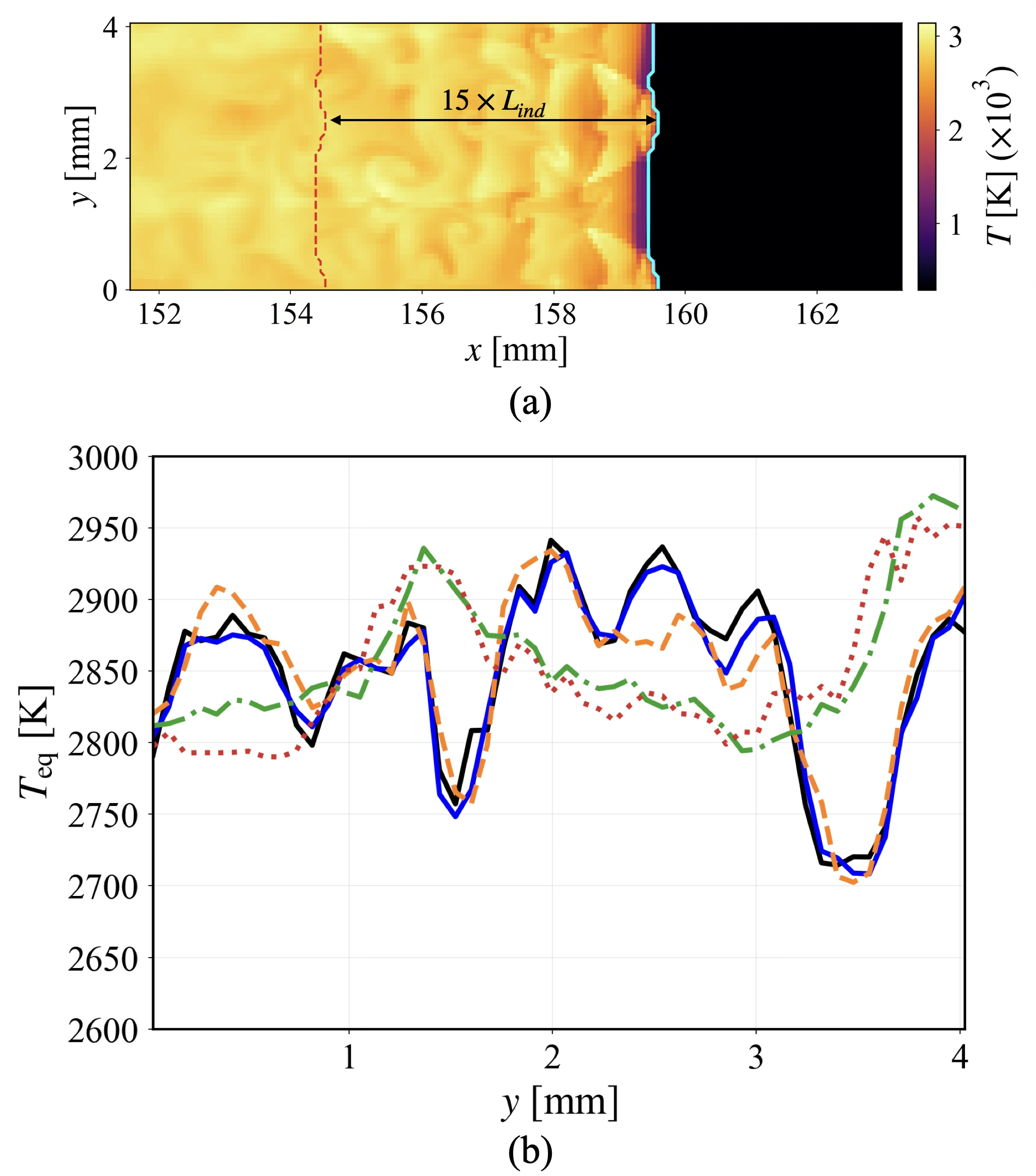}
    \caption{Equilibrium temperature evaluation for the same 2D-Uniform test-set snapshot used in Fig.~\ref{fig:field_predictions_over_eta}. \textbf{(a)} True temperature field, showing the wavefront (\linesample[cyan]{}) and the downstream $15 \times L_{ind}$ location (\linesample[red]{dashed}) at which equilibrium temperature is evaluated. \textbf{(b)} True equilibrium temperature profile along $y$ (\linesample[black]{}) compared with predictions from models with $\eta =1$ (\linesample[blue]{}), $\eta =2$ (\linesample[orange]{dashed}), $\eta =5$ (\linesample[tabgreen]{dash dot}), and $\eta =10$ (\linesample[red]{dotted}).}
    \label{fig:eq_temp}
\end{figure}

The post-shock reconstruction can be evaluated better by analyzing the equilibrium temperature, a quantity of interest (QoI) that indicates reaction completeness and provides a quantitative measure of post-shock reconstruction. The equilibrium temperature profile across $y$ is computed by detecting the wavefront and sampling the temperature at $15$ times the ZND induction length downstream of the wavefront at each $y$-slice in the temperature field, as shown in Fig.~\ref{fig:eq_temp}(a). The resulting temperature profile captures the transverse variation in the post-shock burned gas. True equilibrium temperature profiles and the reconstructed profiles by different models are presented in Fig .~\ref{fig:eq_temp}(b). The oscillatory trend of the temperature profile corresponds to the presence of transverse waves, which can also be observed in the temperature field plot in Fig.~\ref{fig:eq_temp}(a). Models with $\eta \in \{ 1, 2 \}$ agree well with the truth. However, models with $\eta \in \{5, 10\}$ show an averaging-out trend, as in Fig.~\ref{fig:field_predictions_over_eta}, indicating a nearly uniform burned-gas profile with little transverse variation and an early termination of reactions.

An interesting observation in the temperature profile plots is that the model with $\eta=2$ captures the temperature variations, whereas the model with $\eta=5$ does not, confirming that $\eta=2$ is a threshold length-scale factor for reconstructing the post-shock region. This trend matches the field plots in Fig.~\ref{fig:field_predictions_over_eta} -- the downstream region is well captured at $\eta=2$ and averaged out at $\eta=5$.

\begin{figure*}[t]
    \centering
    \setlength{\abovecaptionskip}{3pt}
    \includegraphics[width=0.7\linewidth]{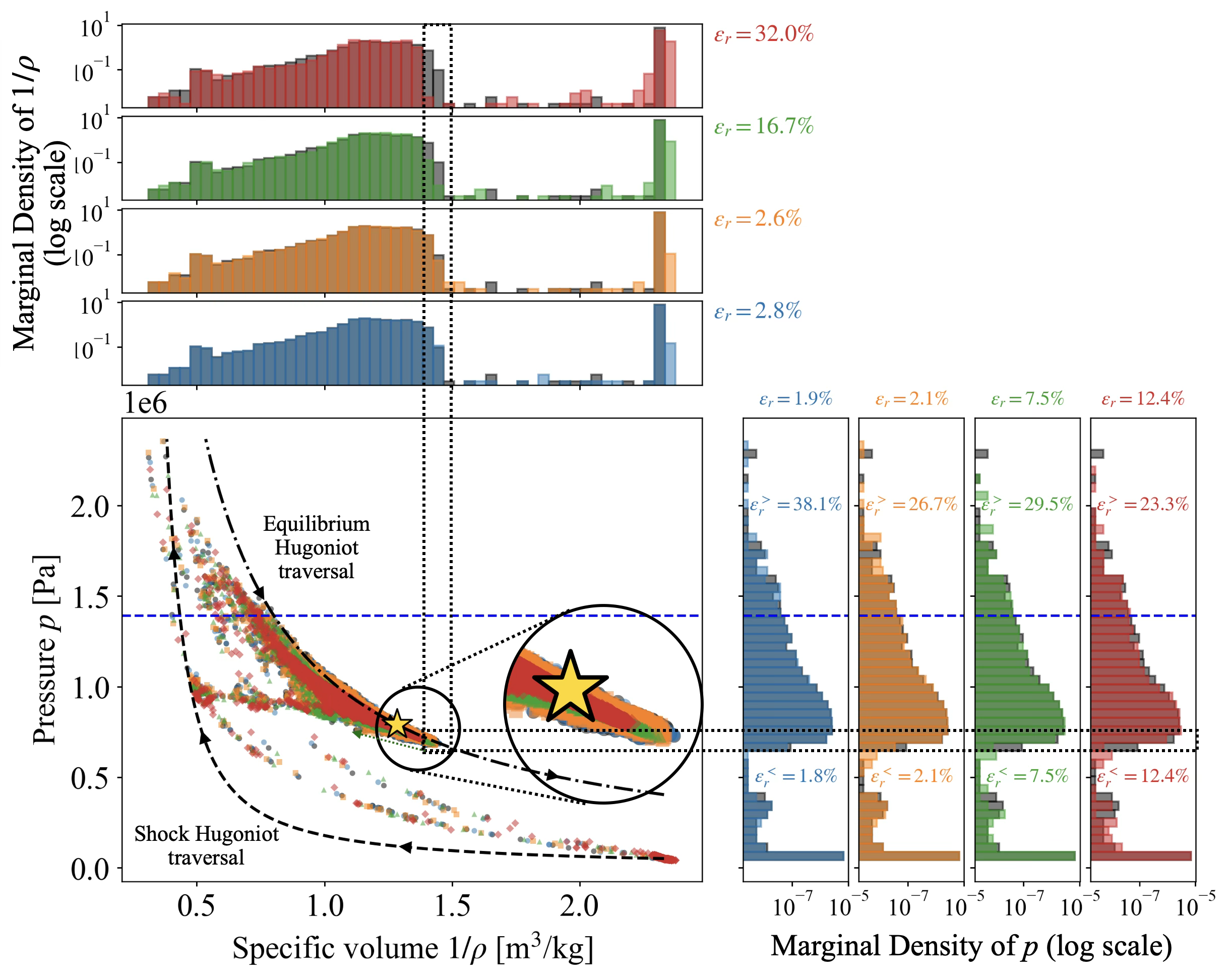}
    \caption{Pressure ($p$) vs.\ specific volume ($1/\rho$) scatter plot for the same 2D-Uniform test-set snapshot used in Fig.~\ref{fig:field_predictions_over_eta}, with marginal density plots of specific volume above the scatter plot and of pressure to the right. The plot includes the truth (\blocksample[black!60]{1.0}) and reconstructions by models with $\eta = 1$ (\blocksample[tabblue]{0.6}), $\eta = 2$ (\blocksample[orange]{0.6}), $\eta = 5$ (\blocksample[tabgreen]{0.6}), and $\eta = 10$ (\blocksample[red]{0.6}). In the scatter plot, \linesample[blue]{dashed} represents the von Neumann pressure, \linesample[black]{dashed} represents the shock Hugoniot curve, \linesample[black]{dash dot} represents the equilibrium Hugoniot curve, and \starsample[yellow] represents the Chapman-Jouguet point in the $p$-$1/\rho$ feature space. $\varepsilon_r$ represents the relative error in the histograms, and $\varepsilon_r^>$ and $\varepsilon_r^<$ represent the relative errors in pressure histogram bins above and below the von Neumann pressure, respectively.}
    \label{fig:pressure_vs_specific_volume_all_models}
\end{figure*}

The effect of the latent graph lengthscale on reaction completeness and wavefront reconstruction can be further understood by studying the pressure ($p$) versus specific volume ($1/\rho$) scatter plot and associated Rankine-Hugoniot curves shown in Fig.~\ref{fig:pressure_vs_specific_volume_all_models}. This data helps to illustrate the flow's trajectory through thermodynamic space, from the ambient pre-shock region to the post-detonation equilibrium state. Alongside the raw data in the scatter plot, marginal PDFs of specific volume and pressure show the distribution of each quantity with different values of $\eta$ compared to the ground truth. From the $p-1/\rho$ plot, as $\eta$ increases, the relative error in the marginals of pressure decreases in the region above the von Neumann pressure (i.e., the maximum pressure in a ZND detonation), where triple points dominate due to transverse-wave interactions and form peak-pressure regions. However, it is emphasized that in this region there is an insufficient number of fluid states to conclude that the models with higher $\eta$ reconstructs the triple point regions better than the models with lower $\eta$. Considering the fact that the order of the relative error in the marginals of pressure in the region above the von Neumann pressure remains the same across $\eta$, and the scarcity of fluid states in this region, the results suggest that movement of $\eta$ from 1 to 10 does not affect the ability of the model to capture these peak-pressure regions. However, the distinction is more pronounced downstream, below the von Neumann pressure, especially near the Chapman-Jouguet (CJ) point, which represents the equilibrium condition in feature space (i.e., complete reactions). From Fig.~\ref{fig:pressure_vs_specific_volume_all_models}, it can be observed that the models with $\eta \in \{1, 2\}$ capture this region well, while the models with $\eta \in \{5, 10\}$ do not fully capture it, as highlighted by a circle in Fig.~\ref{fig:pressure_vs_specific_volume_all_models}. This is analogous to the pattern observed in the equilibrium temperature plot in Fig.~\ref{fig:eq_temp}. The marginal density plots of pressure and specific volume in Fig.~\ref{fig:pressure_vs_specific_volume_all_models} also show that the difference lies in the histogram bins after the CJ condition, i.e., just after a specific volume of $1.28 \, m^3/kg$ and below a pressure of 0.8 MPa (the region highlighted in the $p-1/\rho$ marginal density histograms in Fig.~\ref{fig:pressure_vs_specific_volume_all_models}). Similar trends also appear along the equilibrium Hugoniot curve, where models with $\eta \in \{5, 10\}$ fail to capture the full profile beyond the CJ point (the region highlighted in the $p-1/\rho$ scatter plot in Fig.~\ref{fig:pressure_vs_specific_volume_all_models}).

Overall, the study on $\eta$ shows that $\eta = 2$ can serve as a threshold for the latent graph length scale that strikes a good balance between latent graph resolution and reconstruction fidelity. $\eta = 2$, denoting 0.5 voxels per ZND induction zone length, is considerably coarser than the standard simulation grid requirement of approximately 50-100 cells in the induction zone \cite{barwey2025_chemical_timescales}, yet is able to reconstruct the wavefront and capture the reactions and secondary heat-release regions that contribute to post-shock reconstruction. As such, $\eta = 2$ is used for the extrapolation studies in Sec.~\ref{sec:extrapolation}.

\subsection{Model Performance in Physics Extrapolation Scenarios}
\label{sec:extrapolation}
The results in Sec.~\ref{sec:lengthscale_analysis} revealed the critical role played by the latent graph lengthscale $\eta$ in the reconstruction quality of the detonation wave structure. Specifically, $\eta=2$ (a latent graph resolution of twice the ZND induction length of the H$_2$-air detonation at $p=0.5$~atm) was found to produce an ideal tradeoff between detonation wave structure recovery and resolution reduction in the 2D-Uniform configuration.

Given these findings, the following extrapolation-oriented question is the focus of the present section: \textit{to what degree can graph-based latent representations of simplified cellular detonations (2D Uniform) be transferred to more complex regimes, spanning increased fuel-air stratification reminiscent of real detonation engines (2D Stratified) and richer uniform cellular structures (3D Uniform)?} This is done by extrapolating the $\eta=2$ autoencoder trained on 2D-Uniform data to each of these settings in Sec.~\ref{sec:extrapolation:stratified} and Sec.~\ref{sec:extrapolation:3d} respectively. Such extrapolation is needed to better understand GNN generalizability in out-of-distribution scenarios.

It is emphasized that, per the description of Sec.~\ref{sec:dataset}, snapshots from the three configurations (2D-Uniform, 2D-Stratified, and 3D-Uniform) all span different domain extents (and in the case of 3D, a different spatial dimensionality entirely). Despite this, the analysis described above can still be executed, since message passing layers do not require a fixed connectivity -- rendering domain and dimension-based extrapolation testable.

\subsubsection{Extrapolation to Stratified Detonations}
\label{sec:extrapolation:stratified}

\begin{figure*}[t]
    \centering
    \setlength{\abovecaptionskip}{-10pt}
    \includegraphics[width=1.0\linewidth]{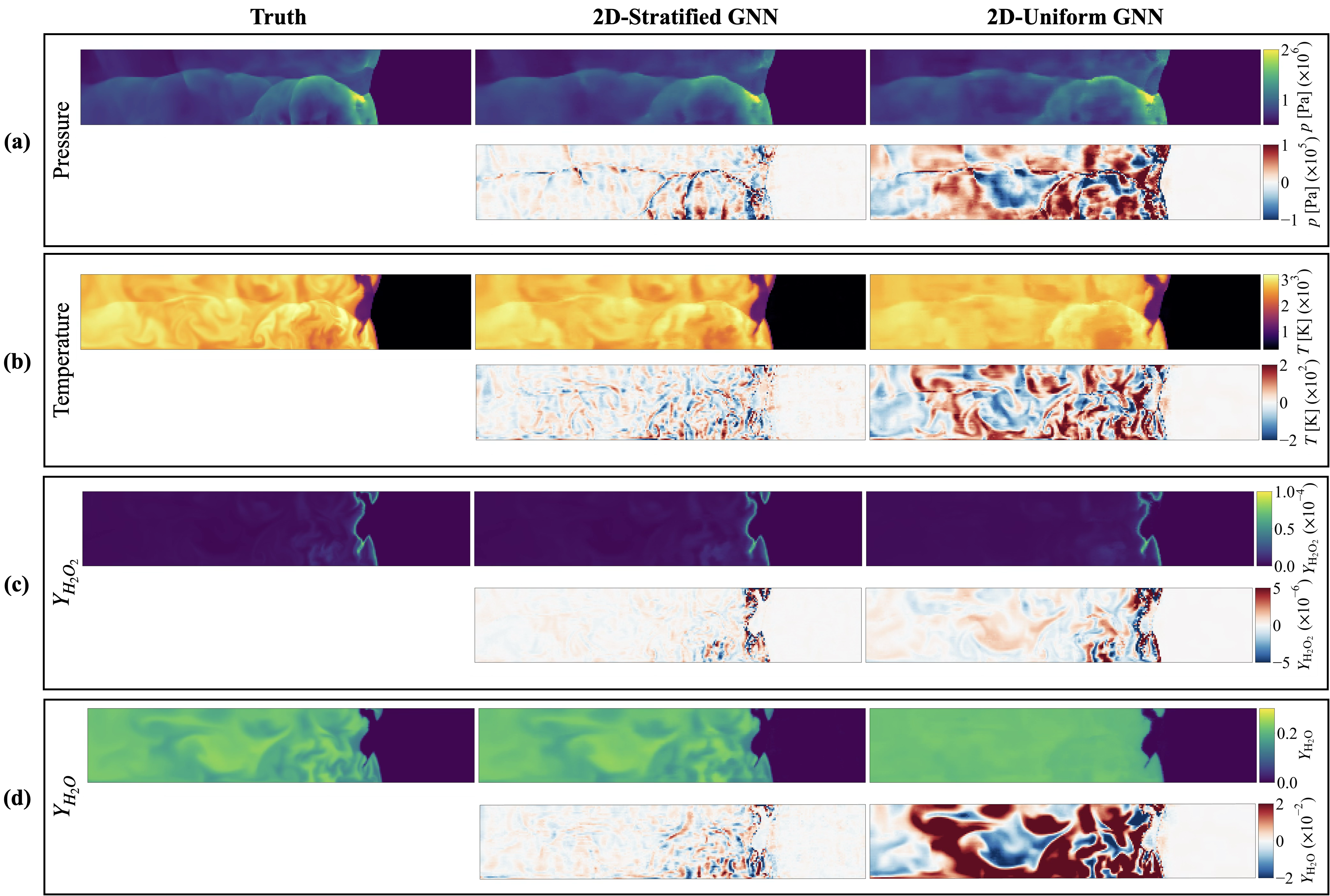}
    \caption{Reconstructions of a 2D-Stratified test-set snapshot for \textbf{(a)} pressure, \textbf{(b)} temperature, \textbf{(c)} $Y_{H_2O_2}$, and \textbf{(d)} $Y_{H_2O}$. Columns show the truth (left), the 2D-Stratified GNN ($\eta=2$) prediction (middle), and the 2D-Uniform GNN ($\eta=2$) prediction (right). In each block, the bottom row shows the corresponding error fields (difference between predicted and true fields).}
    \label{fig:uniform_stratified_field_pred}
\end{figure*}

In this section, the GNN autoencoder trained using 2D-Uniform data (assessed in Sec.~\ref{sec:lengthscale_analysis}) is evaluated on 2D-Stratified data. To disentangle the effect of model-form error versus extrapolation error, the extrapolated 2D-Uniform GNN is compared throughout with a separate GNN \textit{trained} on the 2D-Stratified data at the same settings (i.e., $\eta=2$, $N_{mp}=9$, $H_C=128$). Here, \textit{``2D-Uniform GNN''} and \textit{``2D-Stratified GNN''} refer to the GNNs trained on 2D-Uniform and 2D-Stratified data, respectively.

Before proceeding, it is important to explain the implication of extrapolation in this context. Per the methodology described in Sec.~\ref{sec:methods}, the 2D-Uniform GNN trains message passing layers to encode uniform channel detonations in two spatial dimensions into a latent graph representation characterized by a reduced resolution of $\eta=2$ (two times the ZND induction length). Extrapolation is performed by extracting a 2D-Stratified snapshot and feeding it into the encoder of the 2D-Uniform GNN, which produces a latent graph representation of the 2D-Stratified data as modeled by the message passing layers trained on 2D-Uniform data. This latent graph representation is then projected back onto the fine graph using the decoder of the 2D-Uniform GNN to obtain the final reconstruction.

Figure~\ref{fig:uniform_stratified_field_pred} shows 2D-Stratified (in-distribution) and 2D-Uniform (out-of-distribution) reference  GNN reconstructions of a representative stratified detonation snapshot from the testing set. For the 2D-Stratified GNN predictions, the reconstruction results mirror the in-distribution evaluation trends found for the 2D-Uniform setting in Sec.~\ref{sec:lengthscale_analysis}: the latent representation at the reduced $\eta=2$ resolution sufficiently recovers all of the complexity in the stratified detonation wave structure.

Inspection of the 2D-Uniform GNN predictions in Fig.~\ref{fig:uniform_stratified_field_pred}, on the other hand, reveals the extrapolation capability provided by the GNN autoencoder when moving from uniform to stratified fuel-air mixtures. Interestingly, the 2D-Uniform GNN succeeds in capturing the dominant features, even in the markedly different stratified detonation -- the shape of the wavefront, the location of the triple point, and the associated location of the transverse pressure wave are all captured. In particular, the contorted/wrinkled stratified wave structure is a complexity not observed in 2D-Uniform detonations, yet the 2D-Uniform GNN is able to accurately encode it in its latent representation. This is especially evident in the H$_2$O$_2$ field, in Fig.~\ref{fig:uniform_stratified_field_pred}(c), which shows how the 2D-Uniform GNN is able to recover the intricate structure of the primary reaction zone caused by fuel stratification.

While the wavefront structure is captured, the 2D-Uniform GNN predictions in the downstream reacting region are not as well captured. This is especially apparent in the H$_2$O fields, shown in Fig.~\ref{fig:uniform_stratified_field_pred}(d). Fuel stratification leads to uneven combustion, resulting in unreacted low-temperature fuel pockets \cite{ullman2024_detonation_strutures}. This low-temperature unreacted gas and the high-temperature burned gas undergo turbulent mixing, creating downstream wakes that are more pronounced in the $Y_{H_2O}$ fields, which reflect reaction progress. The latent graph representation of stratified detonations provided by the 2D-Uniform GNN encoder indicates that the transferability of 2D-Uniform detonation physics breaks down in regions of post-shock mixing due to stratification-induced reaction variability.

\begin{figure}[t]
    \centering
    \setlength{\abovecaptionskip}{-10pt}
    \includegraphics[width=1.0\linewidth]{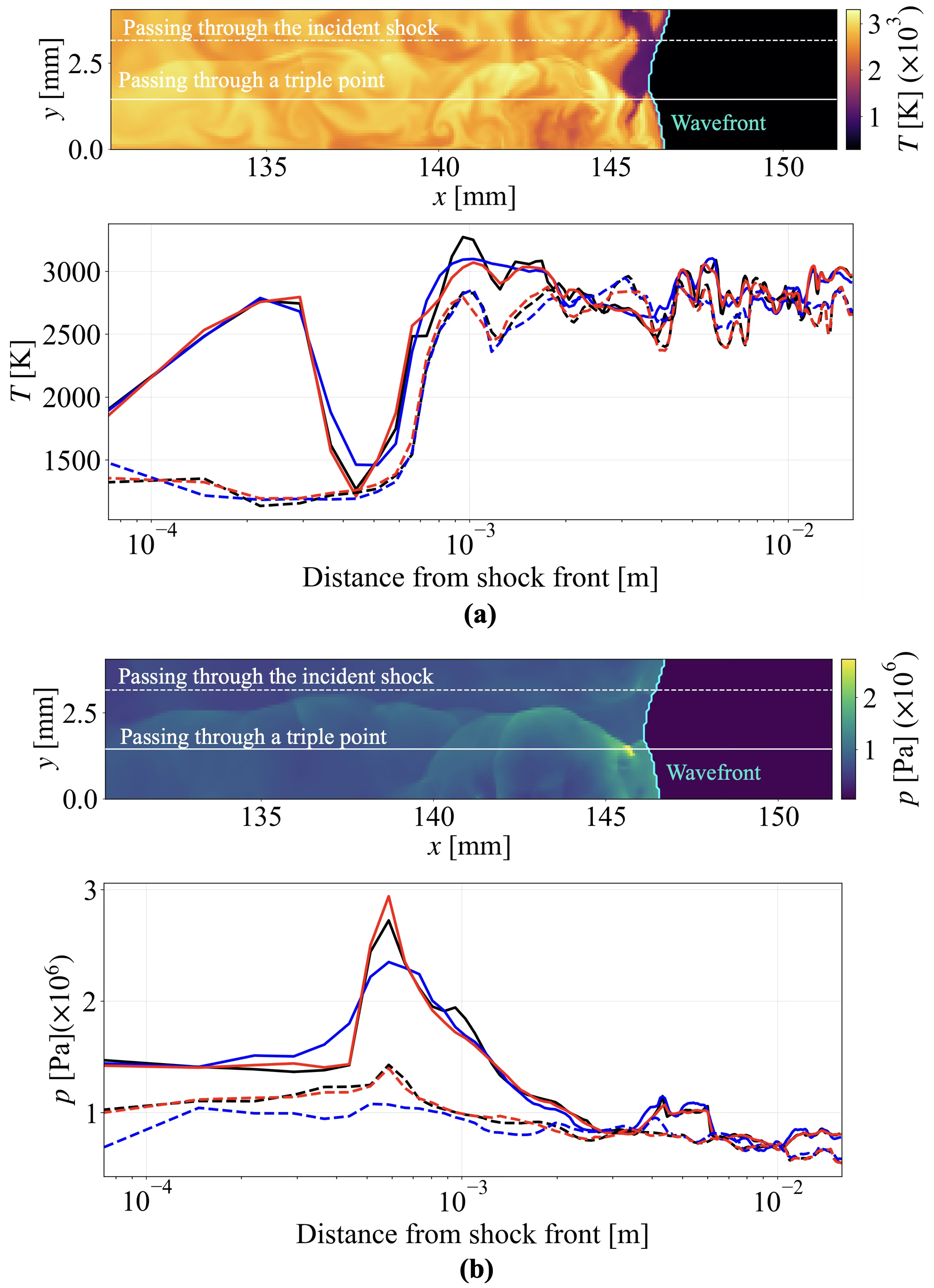}
    \caption{\textbf{(a)} Line plots of one-dimensional slices of temperature, with the full field plot on top showing the locations in $y$ at which the slices were taken. \textbf{(b)} Same as (a), but for pressure. Each line plot depicts the field slices taken along $x$, with the horizontal axis in the shock reference frame (distance from the shock front). Line types depict the type of slice, with \linesample[black]{} representing a slice passing through a triple point and \linesample[black]{dashed} representing a slice passing through the incident shock. Colors represent the true field (\blocksample[black]{}), the field predicted by the 2D-Stratified GNN (\blocksample[red]{}), and the field predicted by the 2D-Uniform GNN (\blocksample[blue]{}).}
    \label{fig:ext_1d_comparison}
\end{figure}

This effect is shown more explicitly in Fig.~\ref{fig:ext_1d_comparison}, which presents one-dimensional temperature (Fig.~\ref{fig:ext_1d_comparison}(a)) and pressure (Fig.~\ref{fig:ext_1d_comparison}(b)) profiles of the stratified detonation extracted along the $x$-direction at two $y$-locations in the shock reference frame: one through the triple point structure and another through the lower-pressure incident shock. These profiles show that (1) the extrapolated temperature profiles (those produced by the 2D-Uniform GNN) are in close agreement with the target until the onset of stratification-induced fluctuations farther downstream, and (2) through the triple point, the peak pressure is underpredicted (with extrapolation exacerbating this effect), but its location relative to the leading shock is correctly captured.

\begin{figure}
    \centering
    \setlength{\abovecaptionskip}{-10pt}
    \includegraphics[width=1.0\linewidth]{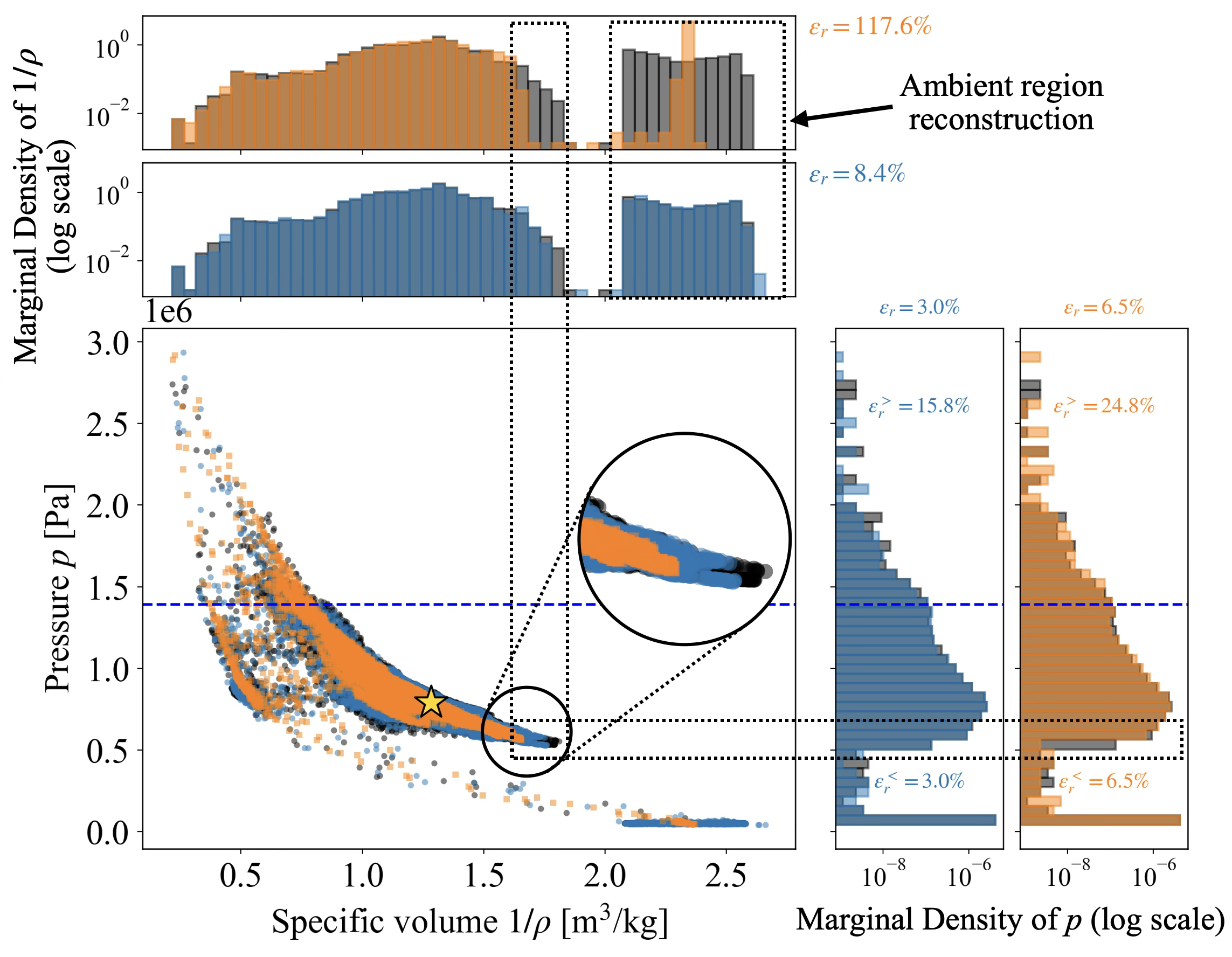}
    \caption{Pressure ($p$) vs.\ specific volume ($1/\rho$) scatter plot for the same 2D-Stratified test-set snapshot used in Figs.~\ref{fig:uniform_stratified_field_pred} and \ref{fig:ext_1d_comparison}, with marginal density plots of specific volume above the scatter plot and of pressure to the right. The plot includes the truth (\blocksample[black!60]{1.0}), the 2D-Stratified GNN prediction (\blocksample[tabblue]{0.6}), and the 2D-Uniform GNN prediction (\blocksample[orange]{0.6}). In the scatter plot, \linesample[blue]{dashed} represents the von Neumann pressure, and \starsample[yellow] represents the Chapman-Jouguet point in the $p$-$1/\rho$ feature space. $\varepsilon_r$ represents the relative error in the histograms, and $\varepsilon_r^>$ and $\varepsilon_r^<$ represent the relative errors in pressure histogram bins above and below the von Neumann pressure, respectively.}
    \label{fig:pressure_vs_specific_volume_us_comparison}
\end{figure}

Inspection of the pressure--specific volume plot in Fig.~\ref{fig:pressure_vs_specific_volume_us_comparison}, analogous to Fig.~\ref{fig:pressure_vs_specific_volume_all_models} for the 2D-Uniform reconstruction, completes the picture. The extrapolation bounds are readily assessed through the marginal probability densities, which again reflect the strong extrapolation capability of the 2D-Uniform GNN in the post-shock, pre-equilibrium region of the stratified detonation. The marginals also reinforce the notable misalignment in the 2D-Uniform GNN reconstructions in the post-CJ state (the region highlighted by a circle in Fig.~\ref{fig:pressure_vs_specific_volume_us_comparison}), which is characterized by an extended equilibrium region beyond the CJ condition when compared to the uniform case in Fig.~\ref{fig:pressure_vs_specific_volume_all_models}. Figure~\ref{fig:pressure_vs_specific_volume_us_comparison} also reveals a key failure point in the 2D-Uniform GNN predictions in the stratified ambient region (highlighted in the specific volume marginal plot as \textit{``Ambient region reconstruction''}): since the fuel-air stratification was never observed during training, the 2D-Uniform GNN cannot be expected to reconstruct the ambient gas variation. More broadly, the reconstruction comparisons between the 2D-Uniform and 2D-Stratified GNNs in the $p$--$1/\rho$ space expose the approximate information overlap \cite{barwey2021_information_overlap}, in the composition space, between 2D-Uniform and 2D-Stratified detonations.

Overall, when faced with a detonation autoencoding task, the GNN consistently prioritizes representing the triple points and the wave profile over the downstream features, even though those downstream features occupy a larger fraction of the coarse and fine graphs (i.e., more node coverage). This behavior is analogous to the trends observed in Sec.~\ref{sec:lengthscale_analysis} when comparing the $\eta=2$ and $\eta=10$ models. There, for in-distribution evaluations, the $\eta=10$ model captured the triple points and leading wave profile (incident shock), but not the post-shock fluctuations. Similarly, in Fig.~\ref{fig:uniform_stratified_field_pred}, the 2D-Uniform GNN captures the stratification-induced effect on the wave profile, but not the coupling between this wave profile and the variation in reaction progress downstream. This points to the fundamental nature of data-based graph representations of detonations: message passing can extrapolate the wavefront shape (i.e., there is enough information in a 2D-Uniform detonation to properly encode the stratified detonation at the wavefront), but not the effect that this structure has on the downstream gas. This limitation is directly connected to the local nature of message passing operations, which here prioritize the detection of stronger, localized discontinuities over the more distributed mixing processes that occur downstream.

\subsubsection{Extrapolation to 3D Detonations}
\label{sec:extrapolation:3d}

The setup is similar to the extrapolation study performed in Sec.~\ref{sec:extrapolation:stratified}, with some key differences. Specifically, extrapolation of the 2D-Uniform GNN on 3D-Uniform data is performed in two settings. The first is one of pure extrapolation: here, the pre-trained 2D-Uniform GNN (the $\eta=2$ model from Sec.~\ref{sec:lengthscale_analysis}) is used to generate a latent representation of the fine-scale graph of a 3D-Uniform snapshot, and then to reconstruct the three-dimensional field from that representation. The second is one of transfer learning: here, the \textit{decoder} of the 2D-Uniform GNN is fine-tuned on a set of 3D-Uniform training snapshots (see Table~\ref{tab:simulation_cases}), and this fine-tuned decoder is used to generate the 3D reconstruction from the same latent graph.

In both settings, the encoder of the 2D-Uniform GNN populates the latent node features of a reduced 3D latent graph at an effective lengthscale (voxel size) of $\eta=2$ (see Sec.~\ref{sec:methods}). The difference lies in the decoder stage, where either the pre-trained (pure extrapolation) or fine-tuned (transfer learning) decoder is used to produce the final reconstruction. The task is therefore the same in both settings: to reconstruct a 3D detonation using encoded information populated by a 2D-Uniform GNN.

\begin{figure*}[t]
    \centering
    \setlength{\abovecaptionskip}{2pt}
    \includegraphics[width=0.8\linewidth]{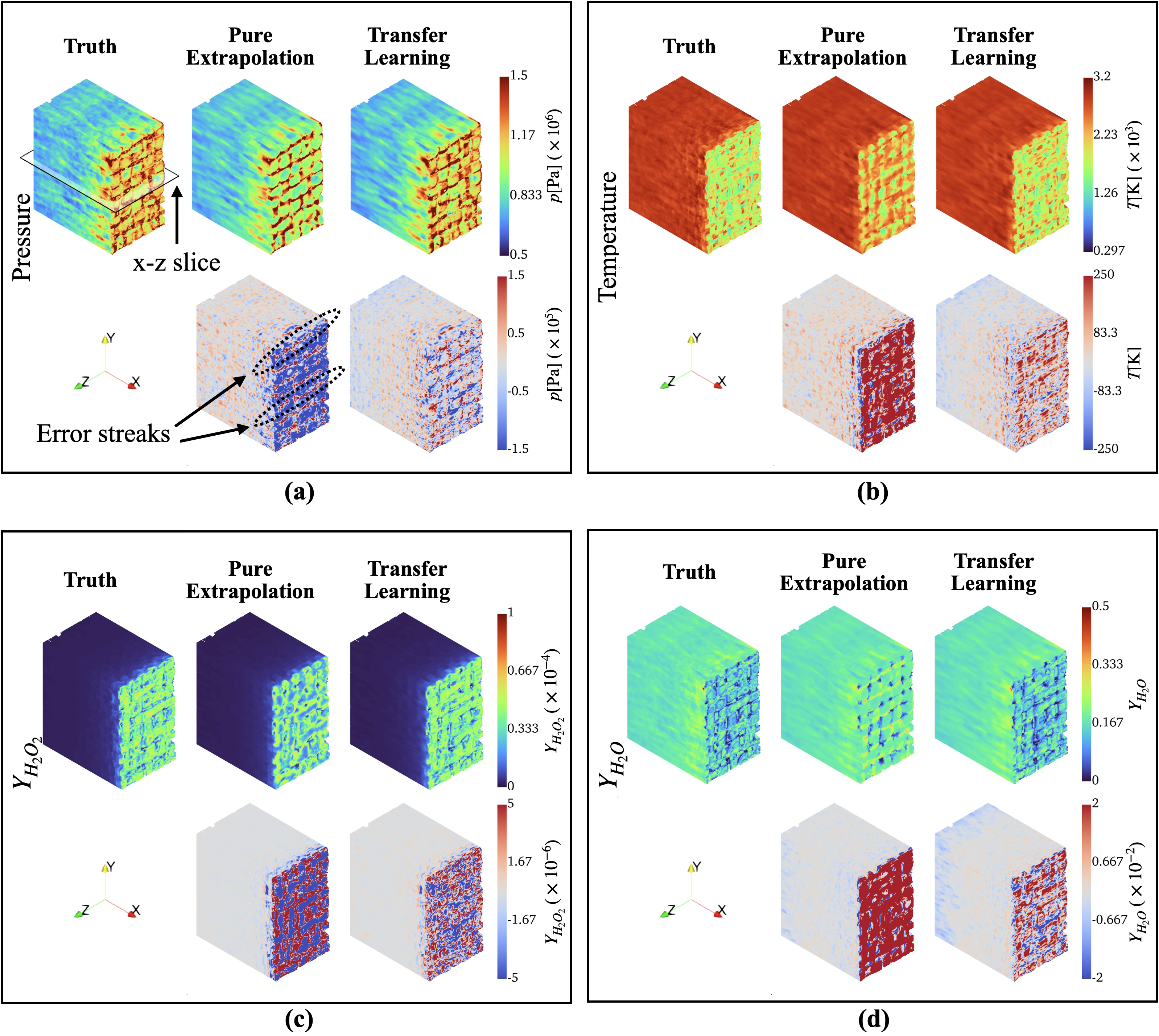}
    \caption{Reconstructions of the 3D-Uniform test-set snapshot for \textbf{(a)} pressure, \textbf{(b)} temperature, \textbf{(c)} $Y_{H_2O_2}$, and \textbf{(d)} $Y_{H_2O}$. Each panel shows the truth (left), the 2D-Uniform GNN prediction with pure extrapolation (middle), and the 2D-Uniform GNN prediction with transfer learning (right), with the bottom row showing the corresponding error fields (difference between predicted and true fields). The $x$-$z$ slice annotation in (a) shows the plane at which the slices in Fig.~\ref{fig:3d_pred_fields_slices} are extracted.}
    \label{fig:3d_pred_fields}
\end{figure*}

\begin{figure}[t]
    \centering
    \setlength{\abovecaptionskip}{-10pt}
    \includegraphics[width=1.0\linewidth]{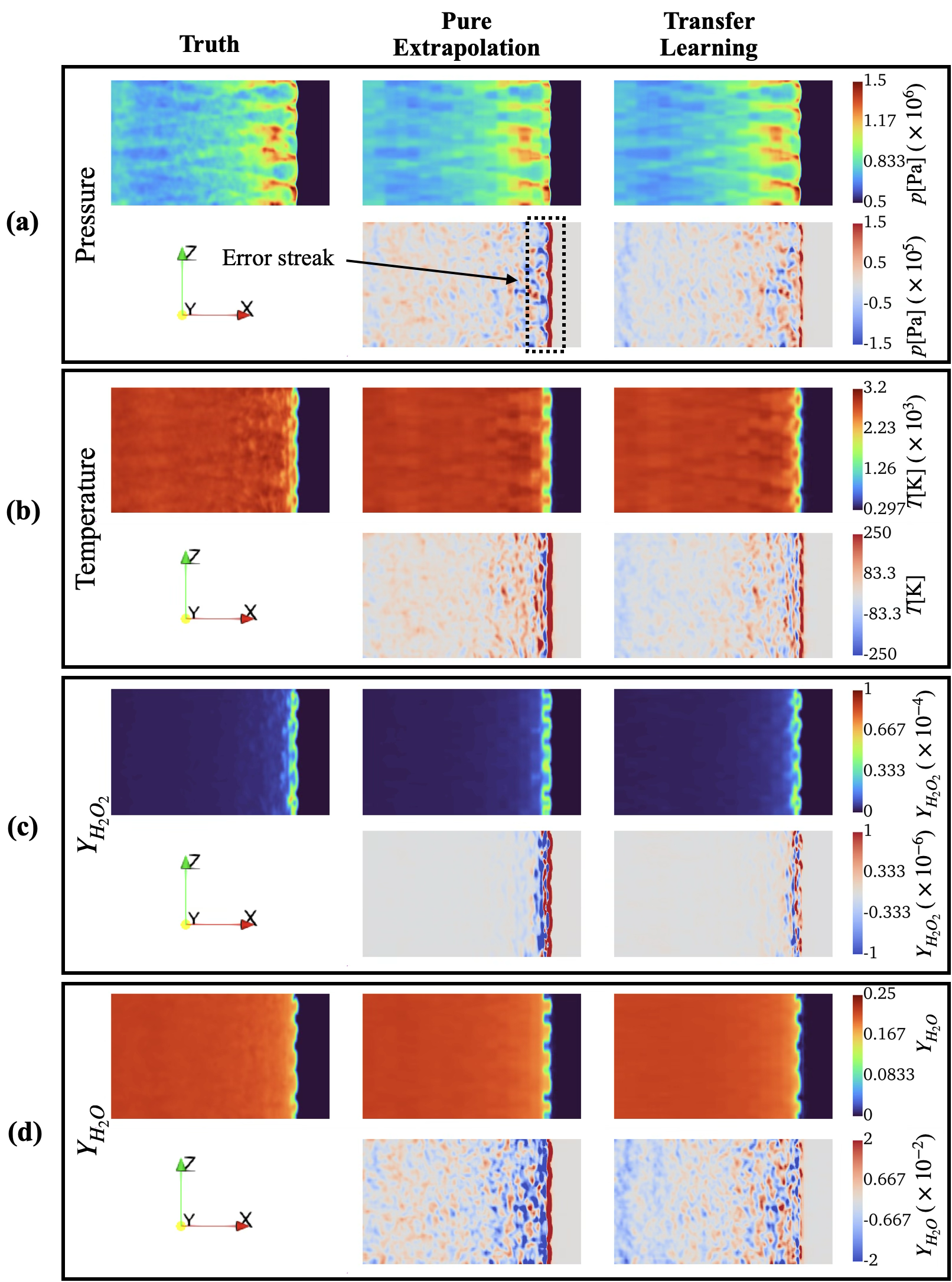}
    \caption{$x$-$z$ slices of the 3D-Uniform test-set snapshot reconstructions in Fig.~\ref{fig:3d_pred_fields}, taken at the plane annotated in Fig.~\ref{fig:3d_pred_fields}(a), for \textbf{(a)} pressure, \textbf{(b)} temperature, \textbf{(c)} $Y_{H_2O_2}$, and \textbf{(d)} $Y_{H_2O}$. Each panel shows the truth (left), the 2D-Uniform GNN prediction with pure extrapolation (middle), and the 2D-Uniform GNN prediction with transfer learning (right), with the bottom row showing the corresponding error fields (difference between predicted and true fields).}
    \label{fig:3d_pred_fields_slices}
\end{figure}

Reconstructions of an unseen 3D-Uniform detonation snapshot by the 2D-Uniform GNN are shown in Figs.~\ref{fig:3d_pred_fields} and \ref{fig:3d_pred_fields_slices}, which visualize the three-dimensional wavefront and the $x-z$ slice at the $y$ mid-point (annotated as ``$x-z$ slice'' in Fig.~\ref{fig:3d_pred_fields}(a)), respectively. Each figure includes the reconstructed detonations and corresponding error fields for both the pre-trained (pure extrapolation) and fine-tuned (transfer learning) 2D-Uniform GNNs.

Inspection of the 3D temperature and pressure fields in Fig.~\ref{fig:3d_pred_fields} shows that the pre-trained 2D-Uniform GNN reconstructs many of the qualitative features of the three-dimensional wave structure. In particular, the locations of the characteristic secondary triple points, caused by three-dimensional transverse wave effects, are recovered in the pure extrapolation reconstructions (this is more apparent in the pressure slices in Fig.~\ref{fig:3d_pred_fields_slices}(a)). However, the pure extrapolation falls short in two respects: (1) pronounced error streaks appear across the $z$-dimension in the pressure error fields (annotated in Figs.~\ref{fig:3d_pred_fields}(a) and \ref{fig:3d_pred_fields_slices}(a)), and (2) the H$_2$O$_2$ and H$_2$O mass fractions are under-represented at the wavefront, indicating incomplete capture of the reaction zone in the 3D structure. The former is reasonable to expect in a 2D-to-3D extrapolation, and the latter is a product of the increase in complexity when moving from 2D to 3D cellular structures. It is nonetheless impressive that a purely 2D model can recover 3D detonations of this quality, which showcases how the GNN is able to capitalize on the overlapping information content between 2D and 3D channel detonations (analogous to what was observed in stratified extrapolations). 

Fine-tuning results in a significant reduction in error across all thermochemical fields. The gain is most noticeable in the H$_2$O$_2$ and H$_2$O mass fraction fields -- the respective 2D slices in Fig.~\ref{fig:3d_pred_fields_slices} show directly how fine-tuning improves the ability to capture the three-dimensional reaction zone at the wavefront, and the shape of the wavefront surface itself. \textit{Importantly, the fine-tuned GNN still operates on the latent representation produced by the original, pre-trained 2D-Uniform GNN encoder.} Its reconstruction capability therefore indicates that 2D GNN-based latent graph representations of detonations have the representational capacity to reconstruct the 3D-Uniform detonation wave structure at an appreciable level of accuracy at the ambient pressure considered ($p=0.5$~atm). However, even with fine-tuning, the 2D-Uniform GNNs exhibit the familiar smearing effect from the stratified extrapolations (Sec.~\ref{sec:extrapolation:stratified}) when reconstructing 3D-Uniform detonations -- most noticeable in the pressure and temperature fields in Fig.~\ref{fig:3d_pred_fields_slices}.

\begin{figure}[t]
    \centering
    \setlength{\abovecaptionskip}{-15pt}
    \includegraphics[width=1.0\linewidth]{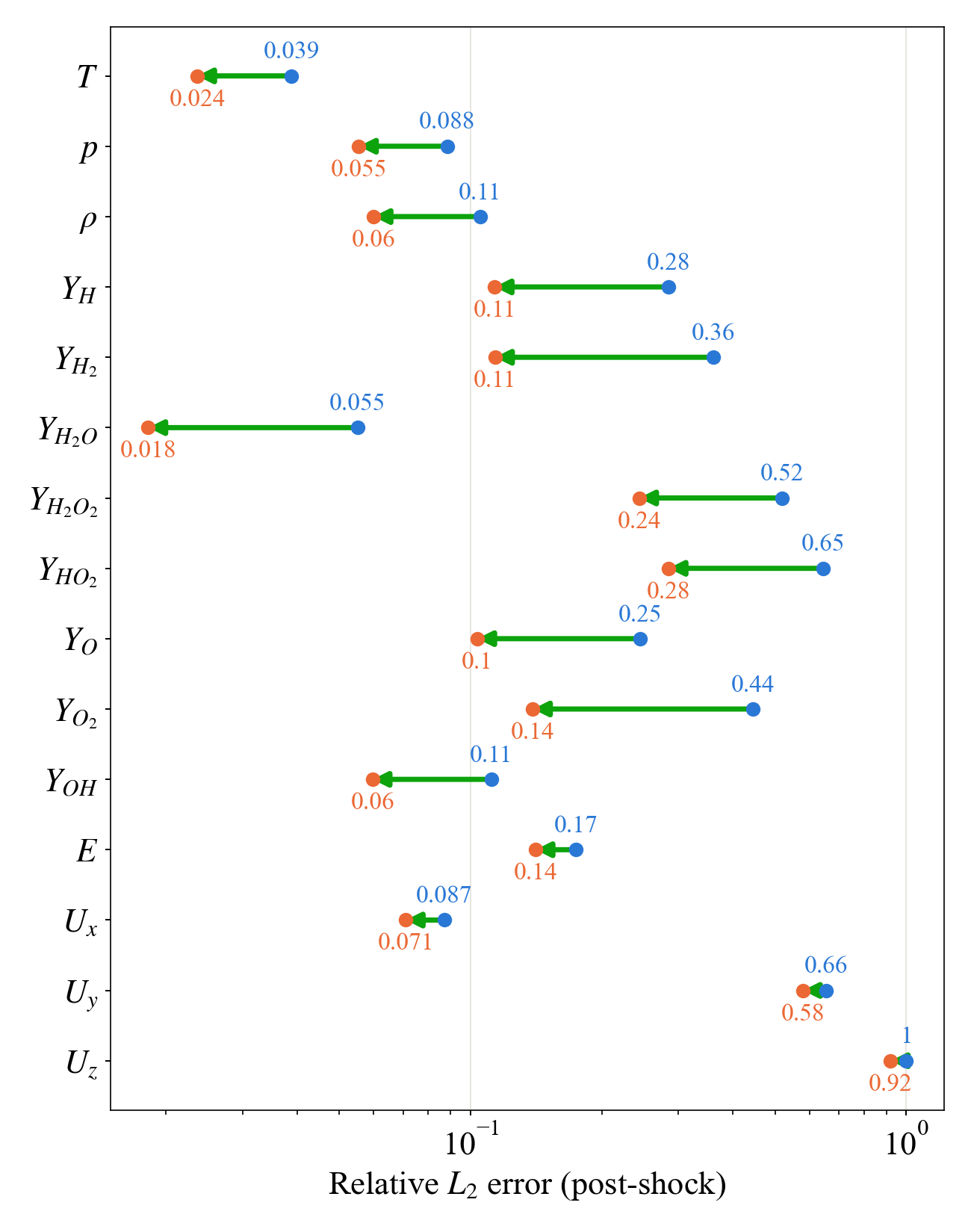}
    \caption{An error plot showing the post-shock relative $L_2$ error (on the $x$-axis) for various fields (on the $y$-axis) in predictions made by the 2D-Uniform GNN with pure extrapolation (\filledcirclesample[tabblue]{}) and with transfer learning (\filledcirclesample[taborange]{}), with \arrowsample[tabgreen]{180} depicting the improvement achieved.} 
    \label{fig:tl_improvements}
\end{figure}

Improvements in individual field variable reconstruction achieved through decoder fine-tuning are shown in Fig.~\ref{fig:tl_improvements}. While fine-tuning improves the prediction of the 3D field variables overall, the effect is more pronounced for the mass fractions of all chemical species, for which the relative errors are reduced by a factor of two or more. This is consistent with the error reductions observed in Figs.~\ref{fig:3d_pred_fields} and \ref{fig:3d_pred_fields_slices}. Note that the error in mass fraction of the combustion byproduct, $Y_{H_2O}$, is much lower than that of the other species. A smaller, but still significant, improvement is observed in temperature, pressure, and density, while only a very small improvement is observed in the flow quantities $U_x$, $U_y$, and $U_z$.

In the reaction zone near the wavefront, the reactants are consumed to form intermediate species, and the reactions complete in the equilibrium region, which lies slightly behind the wavefront, to form final combustion byproducts, i.e., H$_2$O. The lower error in the mass fraction of H$_2$O indicates that the 2D-Uniform GNN captures the field beyond the equilibrium region well, indicating that the 3D field beyond the equilibrium region is similar to that of 2D detonations in uniform channel configurations. This is expected, as the equilibrium thermodynamic state is dictated by conservation of mass, momentum, and energy, independent of the wave's dimensionality. Nevertheless, fine-tuning still improves the predictions in this region. The errors in the reactants (H$_2$ and O$_2$) and intermediate species (H, H$_2$O$_2$, HO$_2$, O, and OH) are significantly reduced. This can be attributed to the improved reconstruction of the reaction zone through transfer learning, a region that is more complex in 3D detonations than in 2D. %

On the other hand, the flow-field quantities $U_y$ and $U_z$ predicted by the 2D model deviate significantly from the true values, especially $U_z$. Since flow in the $z$-direction is never seen by the 2D model, it predicts no variation in $z$. Fine-tuning the decoder with a few 3D snapshots improves the $U_y$ predictions by 12\%, but the improvement in $U_z$ is insignificant. These results indicate that reconstructing an unseen 3D flow field from a 2D latent representation is considerably more difficult for the flow quantities than for the thermochemical quantities -- this is highlighted by the fact that the transverse wave interactions behind the triple points (flow-driven features are significantly more complex in 3D), are not resolved in both pre-trained and fine-tuned reconstructions (as seen in the pressure fields in Fig.~\ref{fig:3d_pred_fields_slices}).

The above results mark the limit of the representational capacity of 2D-GNN-generated latent graphs for 3D detonations. The decoder message passing layers are largely able to take advantage of the physical similarity between 2D and 3D detonations in the thermochemical variables, with notable reconstruction errors found in the reaction zone structure at the wavefront, and in the flow variables. Fine-tuning the decoder significantly helps with the former, but not the latter.

The fact that the thermochemical fields can be reasonably reconstructed in 3D by a fine-tuned 2D GNN decoder points to the advantage of graph-based representations of detonations and their modeling with message passing layers. The implication here is that these layers have sufficient capacity to uncover dimension-wise similarity in the detonation wave structure. While the message passing parameters of the 2D-Uniform GNN \textit{encoder} only have knowledge of the two-dimensional cellular structure, the fine-tuned \textit{decoder} is able to map the resulting latent representations into a plausible three-dimensional cellular structure. Such dimensional-overlap phenomena require further investigation, but these potential correlations, and the associated self-similarity between 2D and 3D detonation wave structures, can be advantageous in modeling 3D detonation evolution via 2D data augmentation. Indeed, it is much easier computationally to produce high-fidelity training data for long-time simulations of 2D detonations than for 3D ones. The present framework provides a methodology to achieve such dimensionality mixing.

\section{Conclusion}
\label{sec:conclusion}

In this work, a graph neural network (GNN)-based autoencoder is leveraged to produce reduced representations of detonation-containing flows. Here, the encoder converts a graph-based representation of an input detonation snapshot (termed the fine graph) into a reduced latent graph-based representation, and the decoder then reconstructs the flow variables on the fine graph using information contained only in this latent graph. Message passing layers are used as the backbone in both steps. The reduction is enabled through a voxel clustering procedure, in which the latent graph nodes can be represented as the nodes of a coarse mesh. The resolution of this coarse mesh is set by the voxel size, which is specified as the quantity $\eta$ times the ZND induction length of the detonation, $L_{ind}$. The so-called latent graph lengthscale $\eta$ provides a direct pathway to assess the resolution reduction in terms of a detonation quantity of interest (the induction length). Through this formulation, this work aimed to understand the degree to which (a) reduced latent graphs can encode the detonation wave structure, and (b) these representations can be extrapolated to more complex detonation configurations. 

To this end, high-fidelity data from three hydrogen-air channel detonation configurations were considered, all at fixed ambient conditions of $p=0.5$~atm and $T=297$~K: (1) \textbf{2D-Uniform}, corresponding to cellular detonations propagating through a uniform, stoichiometric fuel-air mixture in a two-dimensional channel; (2) \textbf{2D-Stratified}, corresponding to the same channel configuration, but with a stratified fuel-air mixture (equivalence ratios ranging from 0.6 to 1.4); and (3) \textbf{3D-Uniform}, the three-dimensional counterpart of 2D-Uniform. The 2D-Uniform configuration was used to assess in detail the effect of the latent graph lengthscale, while the remaining two were used to assess the extrapolation capacity of 2D-Uniform GNN-generated latent graph representations.

The lengthscale analysis in Sec.~\ref{sec:lengthscale_analysis} revealed a clear cutoff at $\eta=2$: latent graphs with a resolution of twice the induction length were able to reconstruct both the wavefront and the post-shock detonation wave structure, despite being significantly coarser than the resolution required in high-fidelity simulations \cite{powers2005_lengthscale,barwey2025_chemical_timescales}. A key finding was that the GNN prioritizes the representation of features at the detonation front (namely, the triple points and the wavefront profile) over the near-equilibrium downstream region characterized by transverse-wave-induced mixing and deflagration. This was most apparent in the pressure--specific volume plots, wherein the models at all tested $\eta$ values sufficiently captured the region above the von Neumann pressure, but only the $\eta \in \{1, 2\}$ models recovered the post-CJ state at a level comparable to the ground truth.

Extrapolation studies were then performed with the $\eta=2$ model, in which latent graph representations generated by the 2D-Uniform GNN were used to reconstruct snapshots in the 2D-Stratified and 3D-Uniform settings. The former was used to test the response to variations in the wave structure due to fuel-air stratification, and the latter constitutes a dimensional extrapolation (2D training and 3D evaluation) to more realistic three-dimensional cellular structures. For 2D-Stratified extrapolations, the 2D-Uniform GNN was able to recover the stratification-induced curvature effect on the wave profile, including the triple point locations (not observed during training), but faced challenges in capturing the coupling between this wave profile and the variation in reaction progress downstream. When applied to 3D-Uniform data, it was found that while the 2D-Uniform GNN was able to recover the thermochemical fields to a reasonable degree (further improved through decoder fine-tuning), the flow variables, particularly the transverse velocity components, were not well recovered, with fine-tuning providing only marginal improvement.

Ultimately, these results reveal a consistent representational hierarchy in GNN-based reductions of detonations: whether the latent graph is coarsened or applied to unseen configurations, retention of the strong features at the detonation front is prioritized over other features. This is directly connected to the local nature of message passing operations, which naturally favor the detection of sharp, localized features, such as shock discontinuities and triple points. A notable finding is that a 2D-Uniform detonation contains sufficient information to encode the wavefront features of both stratified and three-dimensional detonations at the ambient condition considered here. More broadly, alongside reduction, the present results show how GNNs can provide an improved methodology for information overlap extraction and interpretation \cite{barwey2021_information_overlap, procacci_parente_2023_overlap} between complex data distributions -- in this work, this capability was used to effectively ``map'' the salient features of the 2D-Uniform detonation, as contained in the encoder parameters, to reconstruct entirely different detonation wave structures in stratified and three-dimensional detonation configurations.

To conclude, the present findings have direct implications for accelerated simulation of detonation-containing flows through reduced-order modeling, in which the latent graph representations developed here can serve as the reduced state on which detonation dynamics are evolved \cite{rozza2026_graph_ml_surrogate, khodabakshi2026_efficientmodelorderreductiongraph}. This aspect, among others, will be explored in future work.

\section{Acknowledgements}
This research used resources of the Argonne Leadership Computing Facility (ALCF) at Argonne National Laboratory, which is a U.S. Department of Energy Office of Science User Facility operated under Contract No. DEAC02-06CH11357.

\appendix
\section{Global analysis}
\label{appendix:stat_analyses}
\begin{figure}[t]
    \centering
    \setlength{\abovecaptionskip}{-12pt}
    \includegraphics[width=1.0\linewidth]{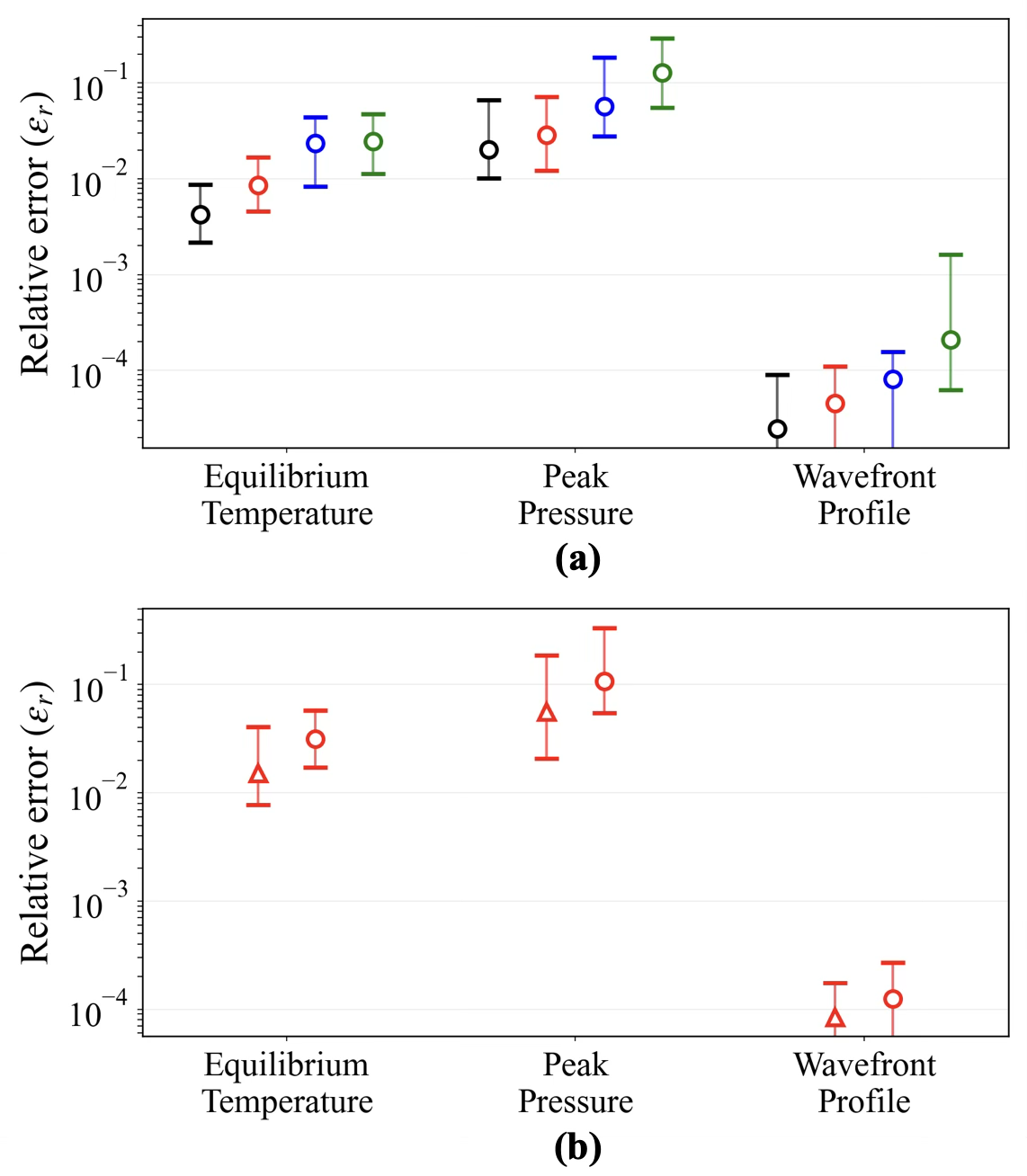}
    \caption{Mean relative error in equilibrium temperature profile, peak pressure, and wavefront profile over a set of test graphs, with error bars indicating minimum/maximum values. \textbf{(a)} Relative errors for 2D-Uniform GNN when applied to 2D-Uniform data, for $\eta = 1$ (\circlesample[black]{}), $\eta = 2$ (\circlesample[red]{}), $\eta = 5$ (\circlesample[blue]{}), and $\eta = 10$ (\circlesample[tabgreen]{}). \textbf{(b)} Relative errors for 2D-Stratified GNN (\trianglesample[red]{}) and 2D-Uniform GNN (\circlesample[red]{}), both with $\eta = 2$, applied to 2D-Stratified data.}
    \label{fig:qoi_error_comparison}
\end{figure}

 In Sec.~\ref{sec:lengthscale_analysis} and Sec.~\ref{sec:extrapolation:stratified}, the analyses were performed on one representative test graph for each 2D-Unifrom and 2D-Stratified case, respectively. In this section, statistical analyses are performed over 2D-Uniform and 2D-Stratified test graphs (see Table~\ref{tab:simulation_cases}) for the following QoIs: equilibrium temperature profile, peak pressure value, and wavefront profile, with relative errors shown in Fig.~\ref{fig:qoi_error_comparison}. The relative errors are calculated using the following equation,

 \begin{equation}
     \varepsilon_{r} = \frac{||q_{\mathrm{pred}} - q_{\mathrm{true}}||_2}{||q_{\mathrm{true}}||_2}, \quad q_{\mathrm{true}}, \, q_{\mathrm{pred}} \in{\mathbb{R}^n},
 \end{equation}

where $ q_{\mathrm{true}}$, and $q_{\mathrm{pred}}$ are the true and predicted values of the QoI, $n$ is the number of mesh/grid points along $y$ for the equilibrium temperature profile and the wavefront profile, and $1$ for the peak pressure, and ``$||.||_2$'' denotes the $L_2$ norm of a vector.

 Figure~\ref{fig:qoi_error_comparison}(a) presents the relative error in QoIs of the 2D-Uniform test graphs predicted by the models of varying latent graph lengthscale $\eta$,  trained with 2D-Uniform data.  The equilibrium temperature error shows a converging trend, with similar error for $\eta = 5$ and $\eta = 10$, which can be justified with the equilibrium temperature profile shown in Fig~\ref{fig:eq_temp}, where an averaging-out trend is observed after $\eta = 2$, resulting from a uniform burned-gas profile. While all the models appear to be capturing the peak pressure location in Fig~\ref{fig:field_predictions_over_eta}(a), from Fig.~\ref{fig:qoi_error_comparison}(a) it is observed that the value of the prediction improves as $\eta$ is lowered, with a swift jump when $\eta$ is increased from $2$ to $5$. Among the three QoIs, the mean relative error in the wavefront profile, extracted by capturing the location in $x$ where pressure exceeds twice the ambient pressure of 0.5 atm, remains consistently low across all models, with minimum values reaching 0 on some test graphs. An averaged out temperature profile and the jump in the peak pressure error above $\eta = 2$ reaffirm the results from Sec.~\ref{sec:lengthscale_analysis} that $\eta = 2$ is a cut-off for capturing the downstream effects while the resolution of the latent graph being significantly coarser than simulation requirements \cite{powers2005_lengthscale, barwey2025_chemical_timescales}.

 The errors in QoIs of the 2D-Stratified test graphs predicted by the 2D-Uniform GNN model and the 2D-Stratified GNN model are presented in Fig.~\ref{fig:qoi_error_comparison}(b). The difference in mean relative error in the equilibrium temperature between the two models results from the difficulty of the 2D-Uniform GNN model in capturing secondary reaction zones arising from a stratified fuel-air mixture. A relatively larger error in the peak pressure predictions by the 2D-Uniform GNN model agrees with Fig.~\ref{fig:ext_1d_comparison}, which shows that the 2D-Unifrom GNN model underpredicts the value of the peak pressure while capturing the peak pressure location. Similar to Sec~\ref{sec:extrapolation:stratified}, a lower wave profile error that is comparable between the 2D-Uniform GNN and 2D-Stratified GNN models suggests that the 2D-Uniform GNN model is able to reconstruct complex wavefront structures in the stratified scenario while facing challenges in capturing the downstream deflagration region, as indicated by the error in the equilibrium temperature.

 Both of these results agree with Sec~\ref{sec:results}: as the complexity of the reconstruction task increases, either by increasing $\eta$ of the model or by reconstructing a graph of a complex scenario, the GNN model prioritizes reconstructing the wavefront structure over other downstream QoIs such as the equilibrium temperature.

{
\footnotesize
\bibliographystyle{elsarticle-num}
\bibliography{references}
}

\end{document}